\documentclass{article}%

\pdfoutput=1

\usepackage{amssymb}
\usepackage{amsfonts}
\usepackage{amsmath}
\usepackage{graphicx}%
\ifx\pdfoutput\relax\let\pdfoutput=\undefined\fi
\newcount\msipdfoutput
\ifx\pdfoutput\undefined\else
\ifcase\pdfoutput\else
\ifx\paperwidth\undefined\else
\ifdim\paperheight=0pt\relax\else\pdfpageheight\paperheight\fi
\ifdim\paperwidth=0pt\relax\else\pdfpagewidth\paperwidth\fi
\fi\fi\fi
\begin{document}

\title{On describing trees and quasi-trees from their leaves}
\author{Bruno Courcelle\\LaBRI, Bordeaux University, France}
\maketitle

\bigskip

\textbf{Abstract }

Generalized trees, we call them \emph{O-trees}, are defined as
\emph{hierarchical} partial orders, \emph{i.e.}, such that the elements larger
than any one are linearly ordered.\ \emph{Quasi-trees} are, roughly speaking,
undirected O-trees.\ For O-trees and quasi-trees, we define relational
structures on their leaves that characterize them up to isomorphism.\ These
structures have characterizations by universal first-order
sentences.\ Furthermore, we consider cases where O-trees and quasi-trees can
be reconstructed from their leaves by \emph{CMSO-transductions}. These
transductions are transformations of relational structures defined by monadic
second-order (MSO) formulas.\ The letter "C" for \emph{counting} refers to the
use of set predicates that count cardinalities of finite sets modulo fixed integers.\ 

O-trees and quasi-trees make it possible to define respectively, the modular
decomposition and the rank-width of a countable graph. Their constructions
from their leaves by transductions of different types apply to
rank-decompositions, and to modular decomposition and to other canonical graph decompositions.

\bigskip

\textbf{Introduction}

In graph theory, a \emph{tree} is a connected graph without root. A
\emph{rooted tree} has a distinguished node called its \emph{root}, from which
one gets a partial order, the \emph{ancestor relation}. A more general notion
of tree has been defined, in particular by Fra\"{\i}ss\'{e} \cite{Fra}
(Section 2.11).\ The notion of a rooted tree is generalized into that of a
\emph{hierarchical} partial order, \emph{i.e.}, such that the set of elements
larger than any one is linearly ordered.\ Such a partial order generalizes the
ancestor relation of a rooted tree.\ A dense linear order, as that of rational
numbers, is then a tree in this extended sense. We call \emph{O-trees} these
generalized trees, by keeping the standard terminology for usual ones, and
\emph{join-trees} those where any two nodes $y,z$ have a least common
"ancestor" $y\sqcup z$, \emph{i.e.}, a least upper-bound with respect to the
associated partial order also called shortly their \emph{join}.

Apart from its intrinsic interest in the theory of relations, the notion of
join-tree permits to define the modular decomposition of a countable graph
\cite{CouDel} by extending the existing notion for finite graphs. Its
formalization in Monadic Second-Order (MSO) logic involves the description of
a join-tree from its leaves forming a set $L$, related by the ternary relation
$R$ such that $R(x,y,z)$ means $x\leq y\sqcup z.$ We call such $(L,R)$ a
\emph{leaf structure}.\ This logical description has been studied in
\cite{Boj,Cam+,CouX} for finite trees, and in \cite{CouDel} for infinite
join-trees. The article \cite{Cam+} shows applications to several canonical
graph decompositions, in particular, the \emph{split} and the \emph{bi-join} decompositions.

In \cite{CouRwd} we also defined \emph{quasi-trees} as, roughly speaking,
undirected join-trees. The motivation was to define the rank-width a countable
graph in such a way that the rank-width of a graph is the least upper bound of
those of its finite induced subgraphs, as for tree-width \cite{Twd}. The
definition of a quasi-tree is based on a ternary relation $B$\ on nodes called
\emph{betweenness} : $B(x,y,z)$ means that $y$ is on the unique path between
$x$ and $z$ in the case of a tree.\ It has been defined and studied in other
cases \cite{Cha+,Chv,Cou20}. We call \emph{leaves} the nodes $y$ that never
occur in the middle of any triple $(x,y,z)$ in $B$. Betweenness is empty on
leaves and is thus useless for describing quasi-trees from their
leaves.\ Hence, we introduce a 4-ary relation on leaves that we call the
\emph{separation relation}:\ $S(x,y,z,u)$ means that $x$ and $y$ on the one
hand and $z$ and $u$ on the other are separated by at least two internal
nodes.\ \ We extend this notion to the induced substructures of quasi-trees
that we call \emph{partial quasi-trees}.\ We represent them topologically as
embeddings in the trees of half-lines that we have already defined in
\cite{Cou21}.

In the present article, we obtain first-order (FO) axiomatizations of the leaf
structures of join-trees and of the separation structures of quasi-trees. We
also examine the constructions of O-trees and quasi-trees by logically
specified transformations of relational structures called \emph{monadic
second-order (MSO) transductions}. We extend to large classes of join-trees a
construction by Bojanczyk \cite{Boj} for finite trees\footnote{A different
construction using a CMSO-transduction is given in \cite{Cam+}.} based on
\emph{counting monadic second-order (CMSO) logic}, the extension of MSO\ logic
that uses set predicates $C_{n}(X)$ expressing that a set $X$\ is finite and
that its cardinality is a multiple of $n$.

The constructions of \cite{CouX} for finite trees and of \cite{CouDel} for
join-trees use auxiliary linear orders and are \emph{order-invariant} which
means that the resulting structures do not depend, up to isomorphism, on the
chosen order.\ That a set has even cardinality is an order-invariant
MSO\ property. Order-invariant MSO\ formulas are more expressive than CMSO
ones \cite{GanRub}.

All countable \emph{leafy} join-trees (such that every internal node is the
join of two leaves) can be constructed from their leaf structures $(L,R)$ by a
fixed order-invariant MSO transduction.\ We extend the constructions of
\cite{Boj,Cam+} to certain leafy join-trees.\ Furthermore, by using
substitutions to leaves in O-trees, we build \emph{CMSO-constructible}%
\ classes of infinite join-trees\emph{ }that can be recontructed from their
leaf structures by CMSO-transductions.

\bigskip

Section 1 reviews O-trees and axiomatizes the leaf structures of
join-trees.\ Section 2\ investigates the constructions of leafy join-trees
from their leaves by CMSO-transductions.\ Section 3 examines the related
questions for quasi-trees.

\section{Order-trees}

All trees and related logical structures will be finite or countable. We will
denote by $\simeq$\ the isomorphism of relational structures and
\emph{u.t.i.}\ to mean \emph{up to isomorphism} when comparing two relational structures.

\subsection{O-trees and join-trees}

Definitions and notation are from \cite{Cou17,Cou21,Cou22,CouDel,Fra}.\ 

\bigskip

\textbf{Definitions 1} : \emph{O-trees}.

(a) A partial order $\leq$ on a set $N$ is \emph{hierarchical} if it satisfies
the \emph{linearity condition,}\ i.e., if for all $x,y,z$ in $N$, $x\leq
y\wedge x\leq z$ implies $y\leq z\vee z\leq y$. If so, $T=(N,\leq)$ is an
\emph{O-forest} and $N$ is its set of \emph{nodes}. If any two nodes have an
upper bound, we call it an \emph{O-tree}.\ The comparability graph of an
O-tree is connected and so is its Gaifman graph\footnote{The \emph{Gaifman
graph }of a relational structure $(D,R)$ has vertex set $D$ and edges between
any two elements occuring in some tuple of $R$.}.

A minimal element is called a \emph{leaf} and $L$ denotes the set of leaves. A
node that is not a leaf is \emph{internal}.\ A maximal element of $N$ is
called the \emph{root} of $T$.

A finite O-tree $(N,\leq)$ is a rooted tree, $N$\ is its set of nodes and
$\leq$ is its \emph{ancestor} relation. We will use for finite and infinite
rooted trees the standard notions of \emph{father} and \emph{son}.

The \emph{covering relation} of a partial order $\leq$ is $<_{c}$ such that
$x<_{c}y$ if and only if $x<y$ and there is no $z$ such that $x<z<y$. If in an
O-tree $(N,\leq)$ we have $x<_{c}y$ we say that $x$ is a \emph{son} of $y$ and
$y$ is the (unique) father of $x$.

(b) If $T=(N,\leq)$ is an O-forest, we denote by $T_{x}$ the O-tree whose set
of nodes is $N_{x}:=\{y\in N\mid y\leq_{N_{x}}x\}$ ordered by the restriction
of $\leq$ to $N_{x}.$ Its root is $x$. We call it a \emph{sub-O-tree} of
$T$.\ More generally, if $M\subseteq N$ is \emph{downwards closed}, that is if
$y\leq x$ and $x\in M$ implies $y\in M$, then $(M,\leq_{M})$ is an O-forest
(or an O-tree) and is also called a \emph{sub-O-forest} (or a
\emph{sub-O-tree}) of $T$.\ 

(c) \ An O-tree $T=(N,\leq)$ is a \emph{join-tree} \cite{CouDel} if every two
nodes $x$ and $y$ have a least upper-bound, called their \emph{join} and
denoted by $x\sqcup y$. If $x,y$ and $z$ are nodes of a join-tree, the joins
$x\sqcup y$, $x\sqcup z$ and $y\sqcup z$ are all equal, or two of them are
equal and larger than the third one. Then, $(x\sqcup y)\sqcup(z\sqcup u)$ is
$x\sqcup y$ or $z\sqcup u$ or $x\sqcup z=y\sqcup z=x\sqcup u=y\sqcup u.$

(d) An O-forest is \emph{leafy} if every internal node is the join of two
leaves. No node can have a unique son.Two incomparable nodes may have no join.
$\ $ \hfill$\square$

\bigskip

For an example, consider the O-tree $T:=(\mathbb{N}\times\{0,1\}\cup
\{a,b\},\leq)$ whose order $\leq$ is defined by $(n,i)\leq(m,0)$ if $m\leq n$
and $i$ is 0 or 1, $a,b\leq(n,0).$ Its leaves are the nodes $a,b$ and $(n,1).$
The root is $(0,0)$.\ It is leafy but $a$ and $b$ have no join.

\bigskip

We studied O-trees in \cite{Cou21,Cou22} by focusing on their monadic
second-order definability and its equivalence with regularity, a notion
derived from the classical theory of infinite rooted trees \cite{FPIT}.

\bigskip

\textbf{Example 2}: \emph{Two O-trees}

We construct a somewhat weird example of an O-tree, to indicate that intuition
may be misleading.

We first define a join-tree.\ Let $A$ and $B$ be disjoint countable dense sets
of nonnegative real numbers such that $0\in B$. We take for $B$ the set of
nonnegative rational numbers and $A\subset\mathbb{R-Q}$, say $A:=(\pi
+\mathbb{Q})\cap]0,+\infty\lbrack.$ We let $N$ be the set of
words\footnote{Words are finite sequences of elements from a possibly infinite
set.} $A^{\ast}\cup A^{\ast}B.$ We define on $N$\ a partial order as follows:

\begin{quote}
$w\leq w^{\prime}:\Longleftrightarrow w=uyv$, $w^{\prime}=ux$ where $u\in
A^{\ast}$, $x,y\in A\cup B$, $v\in N$ and $y\leq x$.
\end{quote}

In other words, we have $w\leq w^{\prime}$ if and only if

\begin{quote}
$w^{\prime}$ is a prefix of $w$ (that is $w=w^{\prime}u$ for some word $u)$,

or $w=uy$, $w^{\prime}=ux$ where $u\in A^{\ast}$, $x,y\in A\cup B$ and $y\leq
x$,

or $w=uyv$, $w^{\prime}=ux$ where $u\in A^{\ast}$, $y\in A,x\in B$, $v\in
A^{\ast}B$ and $y\leq x$.
\end{quote}

Then $T=(N,\leq)$ is a rooted O-tree.\ Its root is the empty word
$\varepsilon$\ and its leaves are the words $u0$ for all $u$ in $A^{\ast}$. An
ascending branch from a leaf $u0$, say $a_{1}a_{2}a_{3}0,$\ to take a
representative example, goes through the following nodes:

\begin{quote}
$u0<a_{1}a_{2}a_{3}b<a_{1}a_{2}a_{3}<a_{1}a_{2}c<a_{1}a_{2}<a_{1}%
d<a_{1}<e<\varepsilon,$
\end{quote}

where $b,c,d,e\in A\cup B$, $b>0$, $c>a_{3}$, $d>a_{2}$, $e>a_{1}$ and there
are infinitely many nodes between any two of $u0,a_{1}a_{2}a_{3}b,a_{1}%
a_{2}a_{3}$ etc.

We have a join-tree because two incomparable nodes $w$ and $w^{\prime}$ are of
the forms $w=uxv$ and $w^{\prime}=uyv^{\prime},$ with $v$ and $v^{\prime}$ non
empty and $x\neq y$.\ Their join is $uz$ where $z:=\max(x,y)\in A.$

Any node $ua$, where $a\in A$, is $u0\sqcup ua0$ but $T$\ is not leafy: the
leaves below $ub$, where $b\in B$ are $u0$ and the $uav0$'s for $a<b$.\ Any
two have upper-bounds of the form $uc$ for $c\in A$, $c<b$, but no least one.

From $T$ as above, we delete all nodes in $A^{\ast}$. The resulting O-tree
$S$\ has the following properties:

\begin{quote}
no two incomparable nodes have a join (they have upper-bounds),

for every internal node $x$, the sub-O-tree $T_{x}$ has infinitely many leaves
and $x$ has a single direction, in $T$ as well as in $T_{x}.$
\end{quote}

The notion of \emph{direction} is defined in \cite{Cou17,CouDel}, Definition
3.2, in both articles.\ Let $x$ be a node that is not a leaf.\ It has a single
direction if for all nodes $y,z<x$ we have $y,z<u$ for some $u<x$.

Any two incomparable nodes $w$ and $w^{\prime}$ as above have infinitely many
upper-bounds $uq$ for the rational numbers $q>\max(x,y)$ but no least one
in\ $A^{\ast}B$.\ \ We have $w,w^{\prime}<uq$, where $x,y<q$ and $w,w^{\prime
}<ur<uq$, for every $r$ in $B$\ such that $\max(x,y)<r<q$. Hence, $uq$ has a
single direction in $S$.

\bigskip

Figures 1 and 2 illustrate this construction. We let $a,b,c,d\in A$ and
$p,q,r,s\in B$ be such that $0<r<q<d<s<c<b<p<a.$

Figure 1 shows some nodes of the O-tree $T$ built from $A$ and $B$.\ The
dotted lines indicate that one has countably many nodes between two shown nodes.\ \ %

\begin{figure}
[ptb]
\begin{center}
\includegraphics[
height=2.514in,
width=3.3252in
]{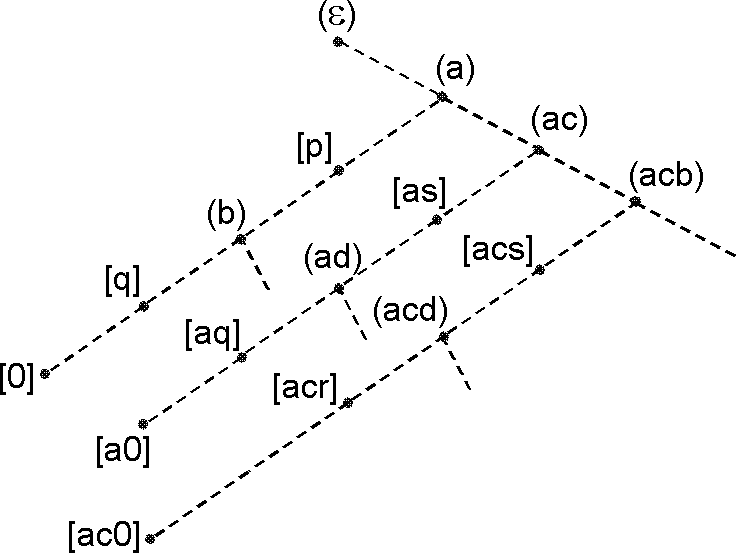}%
\caption{The join-tree $T$ of Example 2.}%
\end{center}
\end{figure}

We denote by $(a)$, $(ac)$, $(acb$) etc. the nodes in $A^{\ast},$ and by
$[p]$, $[aq]$, $[acs]$ etc., the nodes in $A^{\ast}B.$ The root of $T$ is
($\varepsilon$).\ The shown leaves are $[0],[a0]$ and $[ac0].$ In $T$, we
have: $(b)\sqcup(ad)=(b)\sqcup\lbrack as]=[q]\sqcup\lbrack a0]=(a)$ and
$[aq]\sqcup\lbrack acs]=(ac).$

For viewing the corresponding part of the O-tree $S$, just omit the nodes
$(w)$.\ Figure 2 shows the nodes [0]$,[a0],[aq]$ etc.\ The O-tree $S$ has no
root.\ The rightmost branch on Figure 1\ has only nodes in $A^{\ast}$.\ The
three arrows in Figure 2 point to the Dedekind cuts replacing the nodes
$(b),(ad)$ and $(acd)$. \hfill$\square$%

\begin{figure}
[ptb]
\begin{center}
\includegraphics[
height=2.4344in,
width=3.4679in
]{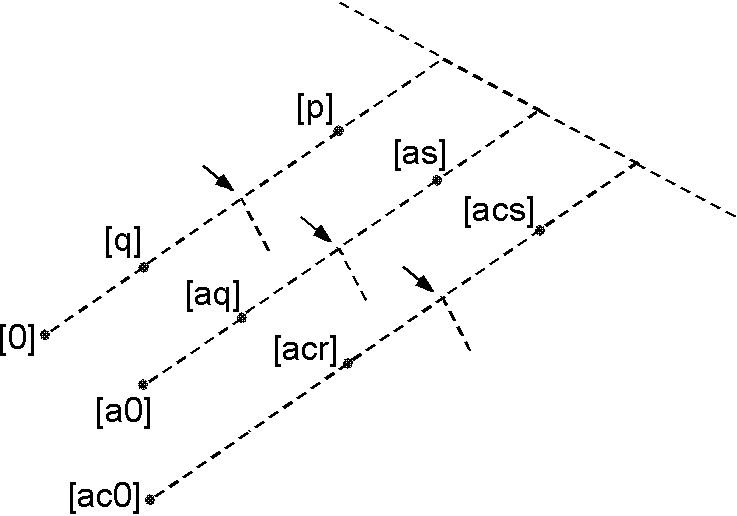}%
\caption{The O-tree $S$ of Example 2.}%
\end{center}
\end{figure}

\bigskip

\bigskip

We let $\mathcal{T}$ be the class of leafy join-trees. This class excludes an
ascending chain like $(\mathbb{Q},\leq)$ and join-trees with nodes having a
unique son or a unique direction (cf.\ Example 2).

\bigskip

Our objective is to describe a leafy join-tree $T=(N,\leq)$ from its
leaves.\ As the order $\leq$ is trivial on leaves, we use instead a ternary relation.\ 

\bigskip

\textbf{Definition 3}\ \cite{CouX,CouXVI,CouDel}\ : \emph{Relations between
leaves.}

For $T$ in $\mathcal{T}$, we let $R$ be the ternary relation on its set of
leaves $L$, defined by $Rxyz:\Longleftrightarrow x\leq y\sqcup z$. We write
$Rxyz$ for $R(x,y,z)$.

We have $x\sqcup y\leq z\sqcup u\Longleftrightarrow Rxzu\wedge$\ $Ryzu$.
Clearly, $Rxyz$ implies $Rxzy$, and $Rxxz$ holds for all $x$ and $z$. If $T$
has only one node $x$, that is thus a leaf and the root, then $Rxxx$
holds.\ If $T$ has one root and two leaves $x$ and $y$, then $Rxxy$ and $Ryxy$
hold.\ Other properties will be given below.

We call $(L,R$) the \emph{leaf structure }of $T$, denoted more precisely by
$(L_{T},R_{T})$.

\bigskip

\textbf{Proposition 4} : If $T$ and $S$ are two trees in $\mathcal{T}$ such
that $(L_{T},R_{T})=(L_{S},R_{S}$), then $T\simeq S.$ \hfill$\square$

\bigskip

We will say that $T$ is \emph{characterized} by its leaf structure
$(L_{T},R_{T}).$

\bigskip

\textbf{Proof}: This is proved for finite trees in Proposition 5.1\ of
\cite{CouX} but the proof works for leafy join-trees. We recall it.

Let $T\in\mathcal{T}$ and $(L,R):=(L_{T},R_{T}).$ Let $j$ be the surjective
mapping : $L\times L\longrightarrow N_{T}$ such that $j(x,y):=x\sqcup_{T}y$.
We have $j(x,y)\leq_{T}j(z,u)$ if and only if $Rxzu\wedge Ryzu$. Hence,
$N_{T}$ is in bijection with $L\times L/\equiv$ where $(x,y)\equiv
(z,u):\Longleftrightarrow Rxzu\wedge Ryzu\wedge Rzxy\wedge Ruxy.$ The quotient
order\footnote{We denote by $[(x,y)]_{\equiv}$ the equivalence class of
$(x,y),$ and by $[(x,y)]$ if\ $\equiv$\ is understood from the context.}
$[(x,y)]_{\equiv}\leq\lbrack(z,u)]_{\equiv}$ is equivalent to $j(x,y)\leq
_{T}j(z,u)$ and is definable from $R$.\ 

If $T=(N_{T},\leq)\in\mathcal{T}$, one can reconstruct $T$ from $N_{T}$ and
$R_{T}$ up to isomorphism.\ If $T$ and $S$ are in $\mathcal{T}$ and
$(L_{T},R_{T})=(L_{S},R_{S}$), then $T\simeq S.$ \hfill$\square$

\bigskip

The following proposition and Theorem 9\ yield an axiomatization of the leaf
structures of the join-trees in $\mathcal{T}$ by a single \emph{universal
FO}\ (\emph{uFO} in short) sentence, that is, with universal quantifications
applied to a quantifier-free formula.\ Furthermore, these formulas will be
written without function symbols.

\bigskip

\textbf{Proposition 5 }: Let $T=(N_{T},\leq)$ be a join-tree.\ The following
properties hold for all leaves $x,y,z,u$ and $v$ where $R:=R_{T}$ :

\begin{quote}
A1:\ $Rxxy,$

A2: $Rxyz\Longrightarrow Rxzy,$

A3: $Rxyy\Longrightarrow x=y,$

A4: $Rxyz\wedge Ryuv\wedge Rzuv\Longrightarrow Rxuv,$

A5: $Rxyz\wedge Rxuv\Longrightarrow(Ryuv\wedge Rzuv)\vee(Ruyz\wedge Rvyz).$
\end{quote}

Furthermore, we have :

\begin{quote}
$x\sqcup_{T}z=y\sqcup_{T}z=x\sqcup_{T}y\Longleftrightarrow R_{T}xyz\wedge
R_{T}yxz\wedge R_{T}zxy.$
\end{quote}

\textbf{Proof :} Easy verifications. Property A3 holds only for leaves
$x,y$.\ The four other properties hold for arbitrary nodes.\ \hfill$\square$

\bigskip

\textbf{Proposition 6} : For every ternary structure $(L,R)$, the following
properties are consequences of the universally quantified formulas A1-A5:

\begin{quote}
A6: $Rxyz\wedge Ruxy\Longrightarrow Ruyz,$

A7: $Rxyz\vee Rzxy$

A8:\ $\lnot Rxyz\Longrightarrow Ryxz\wedge Rzxy,$

A9: $Rxyz\Longrightarrow Ryxz\vee Rzxy.$
\end{quote}

\textbf{Proof :} (1) Property A6 is a consequence of A1 and A4 because:

\begin{quote}
$Ruxy\wedge Rxyz\Longrightarrow Ruxy\wedge Rxyz\wedge Ryyz$ and

$Ruxy\wedge Rxyz\wedge Ryyz\Longrightarrow Ruyz$ by A4.
\end{quote}

(2) Property \ A7\ is a consequence of A1, A2 and A5 : by A1 and A2, we have
$Ryxy\wedge Ryyz$. Then A5\ yields $(Rxyz\wedge Ryyz)\vee(Ryxy\wedge Rzxy).$
By A1 and A2, this reduces to $Rxyz\vee Rzxy.$

(3) We also have $Rxzy\vee Ryxz,$\ hence $Rxyz\vee Ryxz\ $by A2. We
obtain\footnote{Compare it to the last assertion of Proposition 5.} :
$Rxyz\vee(Ryxz\wedge Rzxy)$ that can be rewritten into A8.

(4) From A1-A5,\ we prove A9:\ $Rxyz\Longrightarrow Ryxz\vee Rzxy$ \ by
contradiction; assume this is false for some $x,y$ and $z$.\ We have
$\ Rxyz\wedge\lnot Ryxz\wedge\lnot Rzxy.$ By A8, we get $Rxyz\wedge(Rxyz\wedge
Rzyx)\wedge\lnot Rzxy$ which gives a contradiction by A2. \hfill$\square$

\bigskip

\textbf{Remarks} : (a) Property A5 is not a consequence of A1-A4.\ If one
interprets $Rxyz$ as $x\in\lbrack y,z]\cup\lbrack z,y]$ on real numbers, then
A1-A4\ (whence A6) hold but not A5.\ Properties A7-A9 proved with A5\ do not
hold either.\ 

(b) The following facts have been proved by the prover Z3 \cite{Z3}.

A1-A3, A6 do not imply A4 (cf.\ (1) above).

A1-A3, A7 imply neither A4 nor A5 (cf.\ (2) above).

A1-A4, A9 do not imply A5 (cf.\ (4) above).

However, A1-A4, A7 imply A5.

(c) Let us define

\begin{quote}
$R_{0}xyz:\Longleftrightarrow Rxyz\wedge Ryxz\wedge Rzxy,$ and

$R_{1}xyz:\Longleftrightarrow Rxyz\wedge Ryxz\wedge\lnot Rzxy.$
\end{quote}

Then, by A1,A2 and A8 we have

\begin{quote}
A10: $R_{0}xyz\oplus R_{1}xyz\oplus R_{1}yxz\oplus R_{1}zxy$
\end{quote}

where denotes the exclusive "or".

\bigskip

\textbf{Construction 7 :}\emph{\ From A1-A5 to a join-tree.}

Let $(L,R)$ satisfy properties A1 to A5, universally quantified as
FO\ (first-order) sentences. We construct $T$ in $\mathcal{T}$ such that
$(L,R)\simeq(L_{T},R_{T})$. We use the construction of Proposition 4\ by
assuming only A1-A5, whereas in that construction, we assumed that $(L,R)$ is
$(L_{T},R_{T})$ for some tree $T$ in $\mathcal{T}$.

We define as follows a structure $(L\times L,\sqsubseteq)$ by writing $xy$ for
$(x,y)\in L\times L$ and defining $xy\sqsubseteq zu:\Longleftrightarrow
Rxzu\wedge Ryzu.$ Lemma 8 will show that $\sqsubseteq$ is a quasi-order. Then,
we let $T:=(N,\leq)$ be the corresponding quotient structure. That is, we let

$xy\equiv zu:\Longleftrightarrow xy\sqsubseteq zu\wedge zu\sqsubseteq
xy:\Longleftrightarrow Rxzu\wedge Ryzu\wedge Rzxy\wedge Ruxy,$

we define $N:=L\times L/\equiv$ and $\leq$ as the quotient partial order
$\sqsubseteq/\equiv$.

\bigskip

\textbf{Lemma 8 }: The relation $\sqsubseteq$\ is a quasi-order such that, if
$xy\sqsubseteq zu$ and $xy\sqsubseteq ws,$ then $zu\sqsubseteq ws$ or
$ws\sqsubseteq zu.$

\textbf{Proof : }That $xy\sqsubseteq xy$ (and $xy\sqsubseteq yx$) follows from
A1 and A2.

Property A4\ shows that $Rxzu$ and $zu\sqsubseteq ws$ imply $Rxws$%
.\ Similarly, $Ryzu$ and $zu\sqsubseteq ws$ imply $Ryws$. It follows that
$xy\sqsubseteq zu\sqsubseteq ws$ implies $xy\sqsubseteq ws$, hence that
$\sqsubseteq$\ is transitive and is a quasi-order.

We now prove what will yield the linearity condition for its associated
partial order: if $xy\sqsubseteq zu$ and $xy\sqsubseteq ws$ then, $Rxzu$ and
$Rxws$ hold, and we get $zu\sqsubseteq ws$ or $ws\sqsubseteq zu$ by A5.
\hfill$\square$

\bigskip

\textbf{Theorem 9}\ : The leaf structures of join-trees are the ternary
structures that satisfy A1-A5.

\textbf{Proof :} We use Construction 7 to prove the converse of Proposition
5.\ From $(L,R)$ satisfying A1-A5, we get $T:=(N,\leq)$.\ \textbf{\ }By Lemma
8, $\leq$\ is a partial order and it is hierarchical. Hence, $T$ is an O-forest.\ 

It follows from\footnote{We use A3 only here.} A3 that its leaves are the
elements $[xx]$ for $x\in L$.\ Note that $xx\equiv zu$ if and only if $x=z=u$.

The join of two leaves $[xx]$ and $[yy]$ is $[xy]$, hence $T$ is an O-tree.
Furthermore, $([xx],[yy],[zz])\in R_{T}$ if and only if $Rxyz$ holds because
$Rxyz$ holds if and only if $xx\sqsubseteq yz$.

\bigskip

\emph{Claim}: Any two elements $[xy]$ and $[zu]$ have a join.

\emph{Proof }: If $xy\sqsubseteq zu$ then $[xy]\sqcup\lbrack zu]=[zu].$ If
$zu\sqsubseteq xy$ then $[xy]\sqcup\lbrack zu]=[xy].$

If $Rzxy$ holds, we have $zz\sqsubseteq zu$ and $zz\sqsubseteq xy$, hence
$xy\sqsubseteq zu$ \ or $zu\sqsubseteq xy$ by A5. Similarly if $Ruxy$, $Rxzu$,
or $Ryzu$ holds, then $[xy]\sqcup\lbrack zu]$ is defined and is either
$[xy]\ $ or $[zu].$

If none of these facts holds, from $\lnot Rzxy$, $\lnot Ruxy,$ $\lnot Rxzu$
and $\lnot Ryzu,$ we get by A8 (cf. Proposition 6) and A2 :

\begin{quote}
($Rxzy\wedge Ryzx)\wedge(Rxuy\wedge Ryux)\wedge(Ruxz\wedge Rzxu)\wedge
(Rzyu\wedge Ruyz),$
\end{quote}

which gives $xu\sqsubseteq yz,yu\sqsubseteq xz,xz\sqsubseteq yu$ and
$yz\sqsubseteq xu.$ Hence, $xz\equiv yu$ and $yz\equiv xu.$

We also have $xz\equiv xu$ since $Rxxu\wedge Rzxu$ holds and similarly,
$yz\equiv yu.$ Finally, $xz\equiv$ $xu$ $\equiv yz\equiv yu.$

If $xy\sqsubseteq ws$ and $zu\sqsubseteq ws$, we get $xz\sqsubseteq
ws$,\ hence $[xy]\sqcup\lbrack zu]=[xz].$\ \hfill$\square$

\bigskip

It follows that $T$ is a leafy join-tree.\ \ As said above, its leaves are the
singleton equivalence classes $[xx]$ for $x\in L$.\ \ \hfill$\square$

\bigskip

This construction applies if $L$ is $\{x\}$ or $\{x,y\}$.\ In both cases,
$R$\ is not empty by A1. We obtain respectively a single node $[xx]$ or a
\emph{star} with root $[xy]$ and leaves $[xx]$ and $[yy]$.

We denote by $\tau(L,R)$ the isomorphism class of $T$ constructed as above
from $(L,R)$.\ 

\bigskip

\textbf{Remarks 10 :} (1) From the remarks following Proposition 6, we get
that leaf structures are also axiomatized by A1-A4 and A7.

(2) Some properties of $\tau(L,R)$ are FO-expressible in $(L,R).\ $For
examples, the properties that for $x,y,z,u\in L$, the node $x\sqcup y$ is the
root or that $x\sqcup y$ is a son of $z\sqcup u$, \emph{i.e.,} that $x\sqcup
y<z\sqcup u$ and there is nothing between $x\sqcup y$ and $z\sqcup u.$

(3) The validity of an uFO\ sentence is preserved under taking induced
substructures.\ Hence, if $T=(N,\leq)\in\mathcal{T}$\ and $X\subseteq L_{T}$,
then the (induced) substructure\footnote{If $S=(D,R,...)$ is a relational
structure and $X\subseteq D$,\ then $S[X]:=(X,R[X],...)$ is the
\emph{substructure} of $S$ \emph{induced on} $X$, where $R[X]$ is the
restriction of $R$ to $X$, \emph{i.e.}, the set of tuples in $R$ whose
components are in $X$. This corresponds to the notion of an induced subgraph.
We will only consider substructures of this form.} $(X,R_{T}[X])$ of
$(L_{T},R_{T})$ satisfies properties A1-A5.\ It defines the join tree
$(Y,\leq^{\prime})$ such that $Y=\{x\sqcup y\mid x,y\in X\}\subseteq N$ (this
set contains $X$) and $\leq^{\prime}$\ is the restriction of $\leq$\ to $Y$.

\bigskip

\textbf{Corollary 11}\ : Let $\varphi$ be a first-order sentence over
structures $(N,\leq)$.\ One can construct a first-order sentence $\psi$ such
that for every ternary structure $(L,R)$, the join-tree $\tau(L,R)$ is defined
and $\tau(L,R)\models\varphi$ if and only if $(L,R)\models\psi$.

\textbf{Proof sketch: }It follows from Construction 7\ that the mapping
$(L,R)\longmapsto\tau(L,R)$ is the composition of the following two transformations

\begin{quote}
$(L,R)\longmapsto(L\times L,\sqsubseteq)\longmapsto(L\times L/\equiv
,\sqsubseteq/\equiv)\simeq\tau(L,R)$.
\end{quote}

Since $\equiv$\ is first-order definable in $(L\times L,\sqsubseteq)$, one can
build a sentence $\theta$\ such that, if $\tau(L,R)$ is defined (\emph{i.e.},
satisfies A1-A5), then $\tau(L,R)\models\varphi$ if and only if $(L\times
L,\sqsubseteq)\models\theta$. From $\theta$, one can construct $\psi_{1}%
$\ such that $(L\times L,\sqsubseteq)\models\theta$ if and only if
$(L,R)\models\psi_{1}$.\ Then, one defines $\psi$ as $\psi_{1}\wedge\psi_{2}$
where $\psi_{2}$ is the conjunction of the universally quantified axioms
A1-A5.$\ $ \hfill$\square$

\bigskip

Because of the intermediate step $(L\times L,\sqsubseteq),$\ this construction
does not work for MSO (Monadic Second-Order) sentences $\varphi$\ and $\psi$.
In the next section, we will review \emph{MSO-transductions} and we will
consider cases where $\tau$\ can be defined alternatively by such a
transduction, so that an MSO sentence $\psi$ can be constructed from an
MSO\ sentence $\varphi$\ to satisfy the corollary. In such an alternative
construction, the set of nodes $N$\ is built as a subset of $L\times
\{0,...,k\}$ for some fixed $k$, and not of $L\times L$.

\bigskip

\subsection{Leafy O-forests.}

We let $\mathcal{G}$ be the class of leafy O-forests $T=(N,\leq)$. Our
objective is to extend the results we have for join-trees. For every two
leaves $x,y$, we let $N_{\geq}(x,y):=\{u\in N\mid u\geq x$ and $u\geq y\}$.
Each set $N_{\geq}(x,y)$ is linearly ordered; it is empty if $x$ and $y$ have
no upper bound. If $T$ is an O-tree, none of these sets is empty. Every node
is the join of two leaves, but two nodes may have no join.\ 

\bigskip

We let $R$ be the ternary relation on the set $L$ of leaves defined as follows:

\begin{quote}
$Rxyz:\Longleftrightarrow x\leq N_{\geq}(y,z),$
\end{quote}

that is, $x\leq u$ for every node $u$ in\ $N_{\geq}(y,z)$.\ If $y\sqcup z$ is
defined, then $N_{\geq}(y,z)=N_{\geq}(y\sqcup z)$.\ If $T$ a join-tree, then
$Rxyz\Longleftrightarrow x\leq y\sqcup z$, and this definition is equivalent
to that of Definition 3.\ 

\bigskip

We recall a construction from\ \cite{Cou21}, Section 1.3.\ From an O-forest
$T$, one can construct a join-tree $J(T)$ by adding to it the "missing joins".
Hence, $T$\ embeds into $J(T)$ and $J(T)=T$\ if $T$ is a join-tree.

\bigskip

\textbf{Proposition 12}\ (Proposition 1.4 of \cite{Cou21}) : Let $T=(N,\leq)$
be an O-forest.\ It can be embedded into a join-tree $J(T)$ whose nodes are
the sets $N_{\geq}(y,z)$ for $y,z\in N$.\ Its order is $\supseteq$, the
reverse set inclusion, and $x\in N$\ is mapped to $N_{\geq}(x)\in N_{J(T)}%
$.\ Its leaves are the sets $N_{\geq}(x)$ for $x\in L_{T}$. If $T\in
\mathcal{T}$, then $N_{J(T)}$ is the set of sets $N_{\geq}(y,z)$ for $y,z\in
L_{T}$. If $T$ is not an O-tree, there are $y,z$ such that $N_{\geq}(y,z)$ is
empty, and then, $\emptyset$ is the root of $J(T)$.

\bigskip

\textbf{Proposition 13}\ : If $T$ is an O-forest, then $R_{T}=R_{J(T)}$ and
this relation satisfies A1-A5.

\textbf{Proof} : Let $x,y,z\in L_{T}$.\ We have $N_{\geq}(y)\sqcup
_{J(T)}N_{\geq}(z)=N_{\geq}(y,z).\ $Hence $x\leq N_{\geq}(y,z)$ (\emph{i.e.}
$R_{T}xyz$) if and only if $N_{\geq}(x)\supseteq N_{\geq}(y,z)=N_{\geq
}(y)\sqcup_{J(T)}N_{\geq}(z)$ (\emph{i.e.} ($N_{\geq}(x),N_{\geq}(y),N_{\geq
}(z))\in R_{J(T)}.\ \ $ \hfill$\square$

\bigskip

The O-forest $T$ is isomorphic to a substructure of $J(T)$.\ Since the
formulas A1-A5 are universal (universal quantifications are implicit, cf.
Theorem 9 and Remark 10(3)), their validity is preserved under taking
substructures.\ As they hold in $J(T)$, they hold in $T$. It follows that,
from $(L_{T},R_{T})$ where $T$ is a leafy O-forest, one can define $J(T)$ but
not $T$: clearly, two different O-forests can have the same associated ternary
relation $R$. Note that Property A1\ implies that the Gaifman graph of $(L,R)$
is connected, hence that it cannot hold for an O-forest consisting of several
disjoint O-trees.

\bigskip

A leafy O-forest $T$\ is not characterized by $(L_{T},R_{T})$. We define
$U_{T}$ as the binary relation such that, if $x,y\in L_{T}$, then $U_{T}xy$
\ holds if and only if $x\sqcup_{T}y$ is defined.

\bigskip

\textbf{Proposition 14} : A leafy O-forest $T$ is characterized by (can be
reconstructed from) its associated \emph{extended leaf structure}
$(L_{T},R_{T},U_{T})$.

\textbf{Proof :} From $(L_{T},R_{T})$ one defines $J(T)$ from which one
eliminates the nodes $N_{\geq}(x,y)$ such that $U_{T}xy$ does not hold.
\ \ \ \ \ \ \hfill$\square$

\section{Constructing join-trees by monadic second-order\ transductions}

Monadic second-order transductions are transformations of relational
structures formalized by monadic second-order formulas.

\subsection{Review of definitions and results}

We review informally the definitions, more details can be found in
\cite{CouX,Cou17,CouDel,CouEng}.

A \emph{monadic second-order (MSO) transduction} transforms a relational
structure $S=(D_{S},R_{S},...)$ into $T=(D_{T},R_{T}^{\prime},...)$ by means
of MSO\ formulas collected in a \emph{definition scheme} $\Delta$. To simplify
notation, we let $S$ have a unique relation $R$, and $T$ have a unique binary
relation $R^{\prime}$. The extension to more relations and of different
arities is straightforward.

A definition scheme $\Delta$\ uses set variables $X_{1},...,X_{p}$ intended to
denote subsets of $D_{S}$\ called the \emph{parameters}, and a \emph{copying
constant} $k\geq1$. Then $\Delta=(\chi,\delta_{1},...,\delta_{k},(\theta
_{i,j})_{i,j\in\lbrack k]})$.\ It is a tuple of MSO\ formulas\footnote{For
each positive integer $k$, we denote $\{1,...,k\}$ by $[k]$ \ and
$\{0,1,...,k\}$ by $[0,k]$.} to be evaluated in a structure $S=(D_{S},R_{S}),$
such that

\begin{quote}
$\chi$ has free variables $X_{1},...,X_{p},$

each $\delta_{i}$ has free variables $X_{1},...,X_{p},x,$

each $\theta_{i,j}$ has free variables $X_{1},...,X_{p},x,y.$
\end{quote}

For subsets $X_{1},...,X_{p}$ of $D_{S}$, a structure $T:=\widehat{\Delta
}(S,X_{1},...,X_{p})$ is defined if and only if $S\models\chi(X_{1}%
,...,X_{p}).$ We have\footnote{It may be convenient to have $D_{T}%
\ :=D_{0}\times\{0\}\cup...\cup D_{k}\times\{k\}.\ $\ The copying constant is
then $k+1$.} :

\begin{quote}
$D_{T}\ :=D_{1}\times\{1\}\cup...\cup D_{k}\times\{k\}$ where $D_{i}:=\{x\in
D_{S}\mid S\models\delta_{i}(X_{1},...,X_{p},x)\}$ for each $i$,

$R_{T}^{\prime}:=\{((x,i),(y,j))\mid i,j\in\lbrack k],(x,y)\in D_{i}\times
D_{j}$ and $S\models\theta_{i,j}(X_{1},...,X_{p},$ \ $x,y)\}$.
\end{quote}

The structure $T$ depends on $S$ and $p$ parameters.\ Hence, $S$ has some
image under the multivalued transduction defined by $\Delta$ if and only if
$S\models\overline{\chi}.$ where $\overline{\chi}$ is the sentence $\exists
X_{1},...,X_{p}.\chi.$ However, in our constructions, the structures
constructed from different tuples of parameters will be isomorphic.\ We will
thus write $T=\widehat{\Delta}(S)$ \ where $T$ is defined \emph{u.t.i.} as usual.

\begin{quote}

\end{quote}

The expressive power of MSO\ formulas and MSO\ transductions can be augmented
by using \emph{modulo counting set predicates}: $C_{n}(X)$ expresses that the
set $X$ is finite and that its cardinality is a multiple of $n\geq2$. We refer
to this extension by CMSO.\ The finiteness of a set $X$\ is expressible in
terms of a single set predicate $C_{n}$ (see the proof of Theorem 18).

Yet another extension consists in assuming that the input structure $S$ is
equipped with a linear order\ $\leq$ isomorphic to that of $\mathbb{N}$ if the
structure is infinite. We also assume that the associated transductions are
\emph{order-invariant}, which means that the output structures do not depend,
\emph{u.t.i.}, on the chosen order.\ We denote this extension by $\leq$-MSO
(read \emph{order-invariant MSO}). The property $C_{n}(X)$ is $\leq$-MSO
expressible \cite{CouX}.\ Order-invariant MSO sentences are strictly more
powerful than the CMSO ones on finite structures \cite{GanRub}.

We obtain the notions of CMSO- and \emph{order-invariant} MSO ($\leq
$-MSO)-trans-ductions.\ The latter are more powerful than CMSO transductions,
themselves stronger than the basic ones.

The composition of two MSO (resp. CMSO, $\leq$-MSO) transductions is an MSO
(resp. CMSO, $\leq$-MSO) transduction.

An important result says that if $T=\widehat{\Delta}(S,X_{1},...,X_{p})$, then
an MSO\ (resp. CMSO, $\leq$-MSO) sentence $\varphi$ is valid in $T$ if and
only if some "backwards translated" MSO\ (resp. CMSO, $\leq$-MSO) formula
$\psi(X_{1},...,X_{p})$\ is valid in $S$, where $\widehat{\Delta}$ is,
respectively, an MSO, or CMSO or $\leq$-MSO-transduction.\ See Chapter 7 of
\cite{CouEng}. The corresponding transformation of sentences depends only on
$\Delta.$

Less powerful than the MSO ones are the \emph{FO-transductions} : they are as
above except that the formulas of their definition schemes use neither set
quantifications, nor the set predicates $C_{n}$, nor any auxiliary order. They
can use set variables as parameters.\ The above backwards translation holds
for FO-sentences.

The transformation $(L,R)\longmapsto(L\times L,\sqsubseteq)$ used in Corollary
11 is not an FO-transduction. It could be called a \emph{vectorial
FO-transduction} because the domain of the output structure is a Cartesian
product\footnote{\textbf{Question:} What are the logically defined
transductions that admit a "backwards translation theorem" for FO sentences ?
Among them are the FO-transductions and some vectorial FO-transductions that
remain to be precisely defined.}.

\bigskip

Going back to join-trees, the following theorem collects several
results.\ More will be given in Section 2.4\ below.

\bigskip

\textbf{Theorem 15 : }The following table indicates that, for certain
subclasses of $\mathcal{T}$, the mapping $(L,R)\longmapsto\tau(L,R)\in
\mathcal{T}$, is an MSO-transduction, a CMSO-transduction or an $\leq$-MSO-transduction.\ 

\begin{center}

\begin{tabular}
[c]{|c|c|}\hline
Subclass of $\mathcal{T}$\  & Type of transduction $\tau$\\\hline\hline
finite, degree $\leq k$ & MSO\\\hline
finite & CMSO\\\hline\hline
height $\leq k$ & FO\\\hline
each subtree is finite & CMSO\ \\\hline
$\mathcal{T}$ & $\leq$-MSO\\\hline
\end{tabular}

\end{center}

\textbf{Proofs} : For finite trees of bounded degree, the reference is Theorem
5.6 of \cite{CouX}. Theorem 5.3 of \cite{CouX}\ shows that for general finite
trees, an $\leq$-MSO\ transduction can be used.\ This result is improved in
\cite{Boj,Cam+} with CMSO\ formulas in place of order-invariant ones. We
reproduce below in Theorem 18 the proof of \cite{Boj} and we extend it to
certain infinite join-trees (Theorem 24).\ The case of bounded height is
Proposition 17.\ Proposition 6.1\ of \cite{CouDel} establishes the case of
infinite join-trees. $\ \ $ \hfill$\square$

\bigskip

The following refers to Proposition 14.\ 

\bigskip

\textbf{Corollary 16 }: The mapping $(L_{T},R_{T},U_{T})\mapsto T$ producing
an O-forest $T$ is an $\leq$-MSO transduction.\ It is an MSO- or a
CMSO-transduction in the special cases of Theorem 15.\ 

\subsection{Bounded height}

The \emph{height} of a rooted tree is the least upper-bound of the distances
of the leaves to the root. For each integer $k>0$, we denote by $\mathcal{H}%
_{k}$ the class of\ rooted trees in $\mathcal{T}$\ of height at most $k$, and
by $\mathcal{S}_{k}$ the class of their associated leaf structures.\ Nodes may
have infinite degrees.

\bigskip

\textbf{Proposition 17\ : }The mapping $(L,R)\longmapsto\tau(L,R)$ is a
FO-transduction from $\mathcal{S}_{k}$ to $\mathcal{H}_{k}$.

\textbf{Proof} : Let $(L,R)$ be the leaf structure of some $T$\ in
$\mathcal{T}$. We recall from Theorem 9 that this property is
FO-definable.\ The following properties of leaves $x,y,z,u$ are FO-expressible
in $(L,R)$:

\begin{quote}
(1) $x\sqcup y$ is a son of $z\sqcup u$ in $T$, cf.\ Remark 10,

(2) $\mathrm{depth}(x\sqcup y)=i$ where $\mathrm{depth}(x\sqcup y)$ is the
distance of $x\sqcup y$ to the root; the root is at depth 0.
\end{quote}

The formulas expressing (2) use the one expressing (1).

We fix $k$.\ The leaves are at positive depth at most $k$.\ That
$(L,R)\in\mathcal{S}_{k}$ is FO-expressible by (2) above.

For $0\leq i<k$, we let $D_{i}$ be the set of internal nodes at depth $i$, and
$X_{i}$ be a set of leaves such that, for each $m\in D_{i}$, there is one and
only one leaf $z$ in $X_{i}$ such that $z<m$.\ We let such $z$ be $Rep(m)$:
the leaf \emph{representing} $m$.\ 

An FO-formula $\chi_{i}(X_{i})$ can check that $X_{i}$ is as requested.\ For
each $i=0,...,k-1$, another one, $\rho_{i}(X_{0},...,X_{k-1},x,y,z)$ can check
that $R_{T}zxy$ holds,\ $z\in X_{i}$ and $\mathrm{depth}(x\sqcup y)=i$, so
that $z=Rep(x\sqcup y)$.\ 

The transduction we are consttructing uses parameters $X_{0},...,X_{k-1}$ to
define $D_{T}$ as $X_{0}\times\{0\}\cup X_{1}\times\{1\}\cup...\cup
X_{k-1}\times\{k-1\}\cup L\times\{k\}.$ Here, $L\times\{k\}$\ is the set of
leaves of $T$, and $(u,0)$ is its root, where $X_{0}=\{u\}$ for any leaf $u$.

Its order relation is defined from the following clauses:

\begin{quote}
$(z,k)<(z^{\prime},i)$ where $i<k$ if and only if $Rzxy$ holds for some
$x,y\in L$ such that $\mathrm{depth}(x\sqcup y)=i$, $z^{\prime}\in X_{i}$ (so
that $z^{\prime}=Rep(x\sqcup y$) and $z<x\sqcup y$),

$(z,j)<(z^{\prime},i)$ where $i<j$ if and only if $Rzxy\wedge Rz^{\prime}vw$
holds for some $x,y,v,w\in L$ such that $\mathrm{depth}(x\sqcup y)=i$,
$\mathrm{depth}(v\sqcup w)=j$, $z\in X_{j}$ and $z^{\prime}\in X_{i}$ (so that
$z=Rep(x\sqcup y)$,$z^{\prime}=Rep(v\sqcup w)$ and $v\sqcup w<x\sqcup y$).
\end{quote}

An FO formula $\chi(X_{0},...,X_{k-1})$ can express that the sets
$X_{0},...,X_{k-1}$ satisfy the above conditions.\ We omit further
details.\ \hfill$\square$

\bigskip

All formulas used here are first-order but they use the set variables
$X_{0},...,$ \ $X_{k-1}.$

\subsection{Using CMSO\ transductions}

We extend a construction by Bojanczyck\footnote{A similar more complicated one
is in \ \cite{Cam+}.} \cite{Boj}\ who built a CMSO-transduction that produces
a finite leafy rooted tree from its leaf structure $(L,R)$.\ It yields an
alternative to\ Construction 7\ giving Theorem 9. \ We will prove that this
very construction applies to certain infinite join-trees. In Section 2.4, we
will identify classes of join-trees constructible from their leaves by CMSO transductions.

\bigskip

We recall that in a rooted tree $T$ in $\mathcal{T}$, each\ internal node has
at least two sons. And $Lf_{T}(u)$ denotes the set of leaves below a node $u$.

\bigskip

\textbf{Theorem 18}\ \cite{Boj} : There exists a CMSO\ transduction that
associates with a finite leaf structure $(L,R)$ a finite leafy rooted tree
$T$, such that $R_{T}=R$.\ Furthermore, this transduction defines
$L_{T}=L\times\{1\}$ and $N_{T}-L_{T}\subseteq L\times\{2\}$.

\textbf{Proof idea }: As for the proof of Proposition 17, we will use the
notion of a \emph{representing leaf}. An internal node $u$ of a finite leafy
rooted tree is $x\sqcup_{T}y$ for two leaves $x$ and $y$ and can be identified
with the set $Lf_{T}(u)=Lf_{T}(x\sqcup_{T}y)=\{z\in L_{T}\mid R_{T}zxy\}$. The
construction will select for each such $u=x\sqcup_{T}y$ a representing leaf
$w,$ denoted by $Rep(u)$ that belongs to $Lf_{T}(u)$ and that is
CMSO-definable from $x$ and $y$.\ Then $u$ in $N_{T}-L_{T}$ will be defined as
$(w,2)\in L_{T}\times\{2\}$.\ Each leaf $x$ will be defined as $(x,1)$%
.\ \hfill$\square$

\bigskip

We present the necessary technical definitions and lemmas.

\bigskip

\textbf{Definition 19}\ : \emph{Weights.}

Let $(L,R)$ be the leaf structure of an O-tree $T$.\ We call $\sigma
:L\longrightarrow\{0,1,2\}$ a \emph{weight} mapping.\ We define the
\emph{weight} of a finite set $Y\subseteq L$ as follows: $\sigma(Y)$ is the
sum modulo 3 of the weights $\sigma(x)$ of all $x$ in $Y$. We extend $\sigma$
to internal nodes $u$\ by letting $\sigma(u):=\sigma(Lf_{T}(u))$.

Assume now that $T$\ is finite.\ We say that $\sigma$ is \emph{good} if every
internal node $u$ has a unique son having a maximal value relative to the
order $0<1<2$. This son will be called the\ $\sigma$\emph{-son} of $u$.

\bigskip

\textbf{Lemma 20 (}Claim B5\ of \cite{Boj})\ : For every finite leafy rooted
tree $T$, for every $i$ in $\{0,1,2\}$ and every choice of a "prefered" son
for each internal node, there is a weight $\sigma:L_{T}\longrightarrow
\{0,1,2\}$ such that $\sigma(rt_{T})=i$ and the $\sigma$\emph{-}sons of the
internal nodes are the prefered ones.

\textbf{Proof} : By induction on the size of $T$.\ There is nothing to do if
$r_{T}$ is a leaf.\ Otherwise, let $b_{1},...,b_{p},p\geq2$ be the sons of
$r_{T}$\ and $b_{1}$ be its prefered son.\ Then:

\begin{quote}
if $\sigma(rt_{T})=0,$ we let $\sigma(b_{1}):=2$, $\sigma(b_{2}):=1$ and
$\sigma(b_{i}):=0$ for $i\geq3$,

if $\sigma(rt_{T})=1,$ we let $\sigma(b_{1}):=1$ and $\sigma(b_{i}):=0$ for
$i\geq2$,

if $\sigma(rt_{T})=2,$ we let $\sigma(b_{1}):=2$ and $\sigma(b_{i}):=0$ for
$i\geq2$.
\end{quote}

By the induction hypothesis, $\sigma$ can be defined appropriately on the
subtrees $T_{b_{i}}$ for $i\geq1$.\ \hfill$\square$

\bigskip

\textbf{Proof of Theorem 18 : }Let $(L,R)$ be the leaf structure a finite
leafy rooted tree $T=(N,\leq)$ (not reduced to a single node). Hence,
$\left\vert L\right\vert \geq2$ and $\left\vert N\right\vert \geq3$.

For $i=1,2,$ let $C_{i,3}(U)$ mean that the set $U$ is finite and $|U|=i$ mod
3. It is easy to express $C_{1,3}$ and $C_{2,3}$ in terms\footnote{\ In a
possibly infinite structure, the finiteness of a set $U$ is expressed by
$C_{3}(U)\vee C_{1,3}(U)\vee C_{2,3}(U)$. A similar expression is possible for
each $C_{n}$.} of $C_{3}$.

Let $X_{1}$ and $X_{2}$\ be set variables intended to define the leaves $x$
such that, respectively, $\sigma(x)=1$ and $\sigma(x)=2$. The leaves outside
of these two sets will have by $\sigma$ the value 0.

If $U\subseteq L$, and $\sigma$ is given by disjoint subsets $X_{1}$ and
$X_{2}$ of $L$, we have:

\begin{quote}
$\sigma(U)=i$ if and only if $\left\vert U\cap X_{1}\right\vert +2.\left\vert
U\cap X_{2}\right\vert =i$ mod 3,
\end{quote}

hence, this property is CMSO-expressible by means of the set predicates
$C_{1,3}$ and $C_{2,3}.$

It follows that for every internal node $u$ of $T$\ defined as $x\sqcup_{T}y$,
CMSO\ formulas $\alpha_{i}(X_{1},X_{2},x,y)$ can check whether\ $\sigma(u)=i$
where $\sigma$\ is defined by the disjoint sets $X_{1},X_{2}$ because the set
$Lf_{T}(x\sqcup y)$ is FO-definable in $(L,R)$ from $x$ and $y$.\ Hence, there
is a CMSO\ formula $\beta(X_{1},X_{2},x,y,z,u)$ expressing in $(L,R)$ that
$\sigma$ defined in $T$ by $X_{1},X_{2}$ is good and that $x\sqcup_{T}y$ is
the $\sigma$-son of $z\sqcup_{T}u$. We allow $x=y=x\sqcup_{T}y.$\ A
CMSO-formula $\overline{\beta}(X_{1},X_{2})$ can express that this mapping
$\sigma$ is good.

We need a second good weight mapping $\sigma^{\prime}$, defined similarly by
means of sets of leaves $X_{3}$ and $X_{4}$. Furthermore, we want that each
$\sigma^{\prime}$-son of an internal node differs from the $\sigma$-son.\ A
CMSO\ formula $\gamma(X_{1},X_{2},X_{3},X_{4})$ can express these conditions.
Claerly, $\beta(X_{3},X_{4},x,y,z,u)$ expresses in $(L,R)$ that $\sigma
^{\prime}$ defined by $X_{3},X_{4}$ is good and that $x\sqcup_{T}y$ is the
$\sigma^{\prime}$-son of $z\sqcup_{T}u$.

Let $T=\tau(L,R)=(N,\leq)$.\ We are ready to define a leaf $Rep(x\sqcup_{T}y)$
below any internal node $x\sqcup_{T}y$, intended to represent it, and actually
the subtree $T_{x\sqcup y}$ whose set of leaves is $Lf_{T}(x\sqcup_{T}y)$.

Given $x$ and $y\neq x$, we let $x\sqcup_{T}y,a_{1},...,a_{p}$ be the unique
sequence of nodes of $T$\ such that $a_{1}$ is the $\sigma$-son of
$x\sqcup_{T}y$, $a_{i}$ is the $\sigma^{\prime}$-son of $a_{i-1}$ for $i\geq
2$, until we reach a leaf $a_{p}$ that we define as $Rep(x\sqcup_{T}y).$ The
leaf $a_{p}$ is uniquely defined from $x\sqcup_{T}y$. We have $Rep(x\sqcup
_{T}y)=a_{1}$ if $a_{1}$ is a leaf.

\bigskip

\emph{Claim 1}: If $Rep(x\sqcup_{T}y)=Rep(z\sqcup_{T}u)$, then $x\sqcup
_{T}y=z\sqcup_{T}u.$

\emph{Proof}: Given $a_{p}$, the sequence $a_{p-1},...,a_{1},x\sqcup_{T}y$ is
uniquely defined as each element is the father of the previous one.\ Its stops
as soon as one finds $a_{1}$\ that is a $\sigma$-son, and its father is
$x\sqcup_{T}y.\ \ $\hfill$\square$

\bigskip

\emph{Claim 2} (cf. Theorem 5.3 in \cite{CouX}) : A CMSO\ formula $\rho
(X_{1},X_{2},X_{3},X_{4},x,y,z)$ can express that $z=Rep(x\sqcup_{T}y).$

\emph{Proof: } We have $z=Rep(x\sqcup_{T}y)$ if and only if there exists a set
$\{u_{1},...,u_{p-1}\}$ $\subseteq L$\ \ such that $z\sqcup_{T}u_{1\text{ }}$
is the $\sigma$-son of $x\sqcup_{T}y$, and, for each $i=2,...,p-1$ the
node\ $z\sqcup_{T}u_{i\text{ }}$ is the $\sigma^{\prime}$-son of $z\sqcup
_{T}u_{i-1\text{ }}$, and $z$ is the $\sigma^{\prime}$-son of $z\sqcup
_{T}u_{p-1\text{ }}$. Given a set $U$ intended to be such $\{u_{1}%
,...,u_{p-1}\}$, this condition can be expressed by a CMSO\ formula because
CMSO\ logic can express transitive closure. $\ $\hfill$\square$

\bigskip

Hence, we obtain the tree $\tau(L,R$) with set of leaves $L\times\{1\}$ and
set of internal nodes $Z\times\{2\}$ where $Z$ is the set of leaves that
represent some internal node. $\ $\hfill$\square$

\bigskip

Another CMSO-transduction proving Theorem 18\ is defined in\ \cite{Cam+}: it
uses $C_{2}$ instead of $C_{3}$ and the copying constant is 5. This article
develops applications to several canonical graph decompositions shaped as
trees: the modular decomposition, the split decomposition and the bi-join
decomposition among others.

The proof of Theorem 5.3 of \cite{CouX} uses a auxiliary linear order on
leaves in order to distinguish two sons of each internal node.\ The present
proof uses instead sets $X_{1},...,X_{4}$ from which, however, one cannot
define a linear order of $L$.

\subsection{CMSO-constructible classes}

We will construct subclasses of $\mathcal{T}$ for which some analog of Theorem
18\ holds with CMSO-transductions having possibly more parameters and larger
copying constants.\ Notation is as in Section 2.1.

\bigskip

\textbf{Definitions 21 :} \emph{CMSO-constructible classes of leafy
join-trees.}

(1) A subclass $\mathcal{C}$ of $\mathcal{T}$ is \emph{CMSO-constructible} if

\begin{quote}
(a) the corresponding class of leaf structures is CMSO-definable by some
sentence $\overline{\chi}$ and

(b) there is a CMSO-transduction $\widehat{\Delta}$ that transforms each
ternary structure $(L,R)$ satisfying $\overline{\chi}$ into the corresponding
join-tree $\tau(L,R),$ \emph{u.t.i.}
\end{quote}

(2) A join-tree $T=(N,\leq)$ is \emph{downwards-finite }if its sub-join-trees
$T_{x}$ (cf.\ Definition 1(b)) are finite, for all $x$ in $N$.

(3) We let $\mathcal{D}$ be the subclass of $\mathcal{T}$ consisting of the
finite rooted trees and of the downwards-finite, leafy join-trees without root
(they are necessarly infinite)\footnote{Actually, a downwards-finite, infinite
join-tree cannot have a root.}.

\bigskip

\textbf{Remarks 22\ : }(1) The wording "CMSO-constructible" hides much of the
definition but it is chosen for shortness sake.

(2) Each internal node of a join-tree in $\mathcal{D}$\ has at least two sons.\ 

(3) An example of a join-tree in $\mathcal{D}$\ is\footnote{An infinite
ascending comb without root. It is an infinite \emph{caterpilar}.}
$(\mathbb{N}\times\{0,1\}-\{(0,0)\},\leq)$ whose order $\leq$ is
$(i,j)\leq(k,l)$ if and only if $(i,j)=(k,l)$ or $i\leq k$ and $l=1$. Its
leaves are the nodes $(0,1)$ and $(i,0)$ for all $i\geq1$.\ 

(4) The class $\mathcal{H}_{k}$ is CMSO-constructible (it is even
FO-constructible) by Proposition 17.

\bigskip

\textbf{Lemma 23 }: An infinite leafy join-tree $T$ is in $\mathcal{D}$ if and
only if, for each leaf $x$ and internal node $y>x,$ the interval $[x,y]_{T}$
is finite, and $T$ consists of an infinite ascending branch $x=a_{0}%
,a_{1},a_{2},...,a_{n},...$\ such that for each $i>0$, the node $a_{i}$ has
finitely many sons $a_{i-1},b_{1},...,b_{p}$, where $p\geq1$, and the subtrees
$T_{a_{i}}$ are finite, if and only if the same holds for some leaf $x$.

\textbf{Proof} : Clear from the definitions.\ We have $p>0$\ because $T$ is
leafy.\ Each subtree $T_{b_{j}}$ is finite.\ $\ $ $\ $\hfill$\square$

\bigskip

\textbf{Theorem 24 : }The class $\mathcal{D}$\ is CMSO-constructible by means
of the transduction used for Theorem 18.

\textbf{Proof }: The leaf structures $(L,R)$ of the join-trees in
$\mathcal{D}$ are those that satisfy A1-A5 of Proposition 5\ and the following
CMSO expressible property (that holds trivially if $L$ is finite):

\begin{quote}
for all $x,y\in L$, the set $\{z\in L\mid Rzxy\}$ is finite\footnote{The
finiteness of a set $U$ is expressed by $C_{3}(U)\vee C_{1,3}(U)\vee
C_{2,3}(U)$. \ A similar expression can be based on $C_{n}$ for any $n\geq2.$}.\ 
\end{quote}

For the case where $L$ is infinite, we need only prove the existence of two
weight functions $\sigma$\ and $\sigma^{\prime}$ as in Theorem 18. Let $T$ be
as in Lemma 23. We define $\sigma(a_{i}):=2$\ for each $a_{i}$ and
$\sigma(b_{j}):=0$\ for all its sons $b_{j}$. The values of $\sigma$ on the
nodes of the subtrees $T_{b_{j}}$ exist by Lemma 20\ as these trees are
finite. Hence, the $\sigma$-son of $a_{i}$ is $a_{i-1}$.\ 

We define $\sigma^{\prime}(a_{i}):=0$\ for each $a_{i}$ such that $i$ is even
and 1 if $i$ is odd. For each $i$, we choose one of the sons of $a_{i}$, say
$b_{1}$, to be the $\sigma^{\prime}$-son and we define $\sigma^{\prime}%
(b_{1}):=1$\ if $i$ is odd and $\sigma^{\prime}(b_{1}):=2$ if $i$ is
even.\ All other sons of $a_{i}$ will have by $\sigma^{\prime}$ the weight
0.\ On each subtree $T_{b_{j}}$ we choose $\sigma^{\prime}$-sons to be
different from the already chosen $\sigma$-sons.\ This is possible because
each internal node has at least two sons.\ Weights by $\sigma^{\prime}$\ \ can
be computed on these subtrees by Lemma 20\ again. Then the proof of Theorem
18\ works.\ $\ $\hfill$\square$%

\begin{figure}
[ptb]
\begin{center}
\includegraphics[
height=2.6688in,
width=1.8732in
]{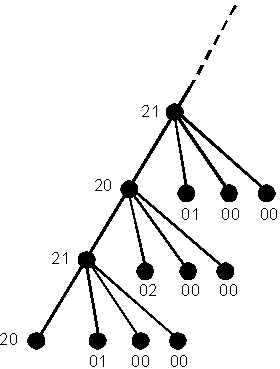}%
\caption{The two good weight functions in the proof of Theorem 24.}%
\end{center}
\end{figure}

Figure 3 illustrates this construction.\ The pair $ij$ at a node $x$ indicates
that $\sigma(x)=i$ and $\sigma^{\prime}(x)=j$.\ 

\bigskip

The CMSO-transduction defined in\ \cite{Cam+} extends to the same class
$\mathcal{D}$ as that of\ \cite{Boj}.\ Our next aim is to construct larger
CMSO-constructible classes of join-trees by combining several known ones. We
use the transductions of \cite{Boj} and, equivalently, of \cite{Cam+}, as
building blocks of more complex transductions.

\bigskip

\textbf{Proposition 25} : If $\mathcal{C}\ $and $\mathcal{C}^{\prime}$ are
CMSO-constructible classes, then, so is their union.

\textbf{Proof }: Let $\Delta$\ be a definition scheme for $\mathcal{C}$\ whose
parameters are $X_{1},...,X_{p}$ and copying constant is $k$, and
$\Delta^{\prime}$\ be similar for $\mathcal{C}^{\prime}$\ with parameters
$Y_{1},...,Y_{q}$ and copying constant $k^{\prime}$. One can construct a
definition scheme $\Theta$ whose parameters are $Z,X_{1},...,X_{p}%
,Y_{1},...,Y_{q}$ and copying constant is $\mathrm{max}(k,k^{\prime})$.\ The
extra parameter $Z$\ decides whether the leaf structure under consideration
should define a joint-tree in $\mathcal{C}$\ (if $Z$ is empty) or in
$\mathcal{C}^{\prime}$ (if $Z$ is not empty). Depending on the case, $\Theta$
uses $\Delta$\ with its parameters $X_{1},...,X_{p}$ or $\Delta^{\prime}%
$\ with its parameters $Y_{1},...,Y_{q}$. We omit the routine details.$\ $%
\hfill$\square$

\bigskip

Every rooted O-tree $T$ is of the form $\ast(T_{1},...,T_{m},...)$ consisting
of a (possibly infinite) disjoint union of O-trees $T_{1},....,T_{m},...$ ,
together with a new root above all of them (denoted by $\ast$). Then $T$ is a
leafy join-tree if and only $T_{1},....,T_{m},...$ are so and their
(unordered) list has at least two O-trees.\ We exclude the trivial case where
$T$ is reduced to a root. However, some of the subtrees $T_{i}$ may be reduced
to a root.

If $\mathcal{C}\subseteq\mathcal{T}$, we let $\mathcal{C}^{\mathcal{+}%
}\subseteq\mathcal{T}$ be the class of join-trees $T=\ast(T_{1},....,T_{m}%
,...)$ such that each $T_{i}$ is in $\mathcal{C}$ or is reduced to a root.
Hence, such $T$ is leafy and belongs to $\mathcal{T}$.

\bigskip

\textbf{Theorem 26 :} If a class $\mathcal{C}$ is CMSO-constructible, then, so
is $\mathcal{C}^{\mathcal{+}}$.\ 

\bigskip

The proof will use \emph{relativizations} of CMSO\ sentences to CMSO-definable
sets (actually the sets of leaves of the $T_{i}$'s).\ This is a standard
notion in logic\footnote{See \emph{e.g.}, \cite{CouEng}, Section 5.2.1.} that
we review.\ 

Let $\varphi(X_{1},...,X_{n},x_{1},...,x_{p})$ be a CMSO formula relative to
relational structures $S=(D,R,...)$ and let $Y$ be a set variable. One can
construct a CMSO\ formula $\varphi\lbrack Y](X_{1},...,X_{n},x_{1},...,x_{p})$
with free variables $Y,X_{1},...,X_{n},x_{1},...,x_{p}$ such that, for every
structure $S$ as above, for all subsets $Y,X_{1},...,X_{n}$ of $D$ and all
$x_{1},...,x_{p}\in D$ we have:

\begin{quote}
$S\models\varphi\lbrack Y](X_{1},...,X_{n},x_{1},...,x_{p})$ if and only if

$x_{1},...,x_{p}\in Y$ and $S[Y]\models\varphi(X_{1}\cap Y,...,X_{n}\cap
Y,x_{1},...,x_{p})$

where $S[Y]$ is the substructure of $S$ induced on $Y$ (cf.\ Section 1.2).\ 
\end{quote}

The construction consists replacing an atomic formula $C_{m}(X_{i})$ by
$C_{m}(X_{i}\cap Y),$ a quantification $\exists x....$ by $\exists x.x\in
Y\wedge...$ and a set quantification $\exists X....$ by $\exists X.X\subseteq
Y\wedge...$ . We omit the details.

\textbf{Proof }: By using Theorem 9 and Remark 10(2), we can write an
FO-sentence $\alpha$\ expressing that a ternary structure $(L,R)$ is the leaf
structure of a rooted join-tree $T$ in $\mathcal{T}$.\ Assuming this,
$T=\ast(T_{1},....,T_{m},...)$ where $T_{1},....,T_{m},...$ belong to
$\mathcal{T}$, have pairwise disjoint sets of nodes $N_{1},....,N_{m}%
,...$\ and sets of leaves $L_{1},....,L_{m},...$, so that $L_{T}=L_{1}%
\cup....\cup L_{m}\cup...$ \ They may have no root. Each relation $R_{T_{i}}$
is the restriction of $R_{T}$\ to $L_{i}$.\ As $T$ is leafy, its root is the
join of two leaves that belong to distinct subtrees $T_{i}$.

Next, we express in $(L,R)$ that\ the $T_{i}$'s are in $\mathcal{C}$. A key
point is to identify by CMSO\ formulas their sets of leaves $L_{i}$.\ For $x$
and $y\in L$, we define $x\sim y:\Longleftrightarrow\exists z.\lnot Rzxy$.

\emph{Claim 1}\ : The relation $\sim$ is an equivalence relation and its
classes are the sets $L_{i}$. Furthermore, $y\sim z\wedge Rxyz\Longrightarrow
x\sim y$.

\emph{Proof }: We need only prove transitivity of $\sim$.\ Assume $x\sim
y\wedge y\sim z.$ We have $\lnot Ruxy\wedge\lnot Rvzy$ for some $u,v$.\ 

We claim that $\lnot Ruxz\vee\lnot Rvxz$ holds, proving $x\sim z$.\ We prove
this from A1-A5 and their consequences A6 and A7.\ A7 yields $Rxyz\vee Rzxy$.
Assume the first.\ If $Rvxz$ holds, then together with\footnote{In such
proofs, we leave implicit the use of A2.} $Rxzy,$ we get $Rvzy$ by A6,
contradicting the hypothesis.\ Assume now $Rzxy$.\ If $Ruxz$ holds, then
$Ruxy$ holds by A6, again a contradiction.\ Hence we have $\lnot Ruxz\vee\lnot
Rvxz$\ and so $x\sim z.$

The condition $\exists z.\lnot Rzxy$\ guarantees that $x\sqcup_{T}y$ is not
the root of $T$, hence that $x$ and $y$ must belong to a same set $L_{i}$.

Assume now $y\sim z\wedge Rxyz.$ Hence, $\lnot Ruyz$ holds for some $u$.\ If
we have $Ruxy$, then with $Rxyz$ and by A6, we have $Ruyz$, contradicting the
choice of $u$. Hence, we have $x\sim y$ and $R$ is stable under $\sim.$
\hfill$\square$

For every equivalence class $M$ of $\sim$, the ternary structure $(M,R[M])$ is
the leaf structure of some leafy\ join-tree that we will denote by $T_{M}%
$.\ These join-trees $T_{M}$ are pairwise disjoint, otherwise, we would have
two classes $M\neq M^{\prime}$ and $x,y\in M,$ $z,u\in M^{\prime}$ such that
$x\sqcup y=z\sqcup u\ $ in $T$, hence $Rzxy$ so that $z\in M$ by Claim1, hence
a contradiction.\ They are the join-trees $T_{1},...,T_{m},...$

We let $\Delta=(\chi,\delta_{1},...,\delta_{k},(\theta_{i,j})_{i,j\in\lbrack
k]})$ be a definition scheme for $\mathcal{C}$, with associated CMSO sentence
$\overline{\chi},$ expressing that $\widehat{\Delta}(L,R)$ is defined.\ 

\emph{Claim 2} : A CMSO\ sentence $\mu$\ can express in $(L,R)$ that each
$T_{M}$\ is in $\mathcal{C}$.

\emph{Proof :} An FO\ formula $\eta(Y)$ can express in $(L,R)$ that a set
$Y\subseteq L$ is an equivalence class of $\sim$. Then, a CMSO\ sentence can
express that for each equivalence class $M$, we have $(M,R[M])\models
\overline{\chi}.\ $ \hfill$\square$

Hence, together with $\alpha$, we have a CMSO\ sentence expressing that
$(L,R)$\ is the leaf structure of some leafy joint-tree $T$ in $\mathcal{C}%
^{\mathcal{+}}.$

\bigskip

To construct $T$, we will define $\Theta=(\chi^{\prime},\delta_{0}^{\prime
},...,\delta_{k}^{\prime},(\theta_{i,j}^{\prime})_{i,j\in\lbrack0,k]}))$ such
that $N_{T}=\{u\}\times\{0\}\cup U_{1}\times\{1\}\cup...\cup U_{k}\times\{k\}$
where $u$ is an arbitrary leaf and $(u,0)$ will be the root of $T$. We will
use the parameters $X_{1},...,X_{p}$ from $\Delta$\ and an additional one
$X_{0}$, intended to hold $u$.\ 

To be convenient, a $p$-tuple ($X_{1},...,X_{p})$ must verify :

\begin{quote}
for each equivalence class $M$, we have $(M,R[M])\models\chi(X_{M,1}%
,...,X_{M,p})$ where $X_{M,i}:=X_{i}\cap M$ for each $i$.
\end{quote}

From such ($X_{1},...,X_{p})$, we define, for each $M$ and $i\in\lbrack k]:$

\begin{quote}
$U_{M,i}:=\{x\in M\mid(M,R[M])\models\delta_{i}(X_{M,1},...,X_{M,p},x)\}.$
\end{quote}

It follows that $N_{T_{M}}=U_{M,1}\times\{1\}\cup...\cup U_{M,k}\times\{k\}$.
To obtain $N_{T}$ as desired, we define $U_{i}$ as the union of the sets
$U_{M,i}$ for all $M$.

Finally, we define $\leq_{T}$ as follows, for $(x,i)\leq(y,j)$ in $N_{T}$:

\begin{quote}
$(x,i)\leq(y,j)$ if and only if

either $(y,j)=(u,0)$,

or $x$ and $y$ are in some $M$ such that $(M,R[M])\models\theta_{i,j}%
(X_{M,1},...,X_{M,p},x,y).$
\end{quote}

From this description, one can (one could) write the formulas\ $\chi^{\prime
},\delta_{0}^{\prime},...,\delta_{k}^{\prime},\theta_{i,j}^{\prime}$ to form
$\Theta.\ $ \hfill$\square$

\bigskip

This proof is a useful preparation for a similar more complicated construction.

\bigskip

\textbf{Definition 27 : }\emph{Substitutions to leaves.}

(1) Let $T=(N_{T},\leq_{T})$ be an O-tree.\ For each $x\in L_{T}$\ , let
$U_{x}=(N_{x},\leq_{x})$ be an O-tree (it may be reduced to a single node).
Since our constructions yield objects \emph{u.t.i.}, we can take them pairwise
disjoint and disjoint from $T$. We let $S=T[U_{x}/x;x\in L_{T}]$ be the
partial order $(N_{S},\leq_{S})$ such that :

$N_{S}$ is the union of $N_{T}-L_{T}$\ and the sets $N_{x}$,

and $u\leq_{S}v$ if and only if

\begin{quote}
either $u,v\in N_{T}-L_{T}$ and $u\leq_{T}v,$

or $u,v\in N_{x}$, $x\in L_{T}$, and $u\leq_{x}v,$

or $u\in N_{x}$, $x\in L_{T}$, $v\in N_{T}-L_{T}$ and $x\leq_{T}v.$
\end{quote}

(2) A small extension of this definition will be useful: we assume that each
$x\in L_{T}$\ is also a leaf of $U_{x}$. There is nothing to change in the
definitions (in the second case above, we may have $u=x$).\ In this case
$L_{T}\subseteq N_{S}$ as each $x\in L_{T}$\ belongs to $N_{x}$.\ In
particular, if $U_{x}=\{x\}$, where $x\in L_{T}$, there is nothing to
substituate to $x$. \hfill$\square$

\bigskip

With these hypotheses and notation:

\textbf{Lemma 28 }: (1) $S=(N_{S},\leq_{S})$ is an O-tree.

(2) If $T$ and the $U_{x}$'s are leafy join-trees, then $S$ is a leafy join-tree.

\textbf{Proof: }(1) is straightforward.

(2) From the definitions, we observe that the joins in $S$ are as follows:

\begin{quote}
if $u,v\in N-L_{T}$ then $u\sqcup_{S}v=u\sqcup_{T}v$,

if $u,v\in N_{x}$, $x\in L_{T}$, then $u\sqcup_{S}v=u\sqcup_{x}v,$

if $u\in N_{x}$, $x\in L_{T}$, $v\in N-L_{T}$ then $u\sqcup_{S}v=x\sqcup
_{T}v,$

if $u\in N_{x}$, $v\in N_{y},x,y\in L_{T}$ and $y\neq x$ then $u\sqcup
_{S}v=x\sqcup_{T}y.$
\end{quote}

Hence, $S$ is a join-tree.\ It is leafy by the above observations.$\ $%
\hfill$\square$

\bigskip

\textbf{Definition 29\ \ :}\emph{ Contracting subtrees.}

(1) Let $S=(N_{S},\leq_{S})$ be a leafy join-tree. A node $x$ is \emph{finite}
if the sub-join-tree $S_{x}$ is finite. Since $S$ is leafy, a subtree $S_{x}$
is finite if it has finitely many leaves.

If $z$ is a leaf, then $Fin(z)\subseteq N_{\geq}(z)$, we denote the set of
finite nodes above $z$ or equal to it. Either $Fin(z)$ has a maximal element
$y$, or there is an infinite ascending branch $z=a_{0},a_{1},a_{2}%
,...,a_{n},...$\ of finite nodes where each $a_{i}$, for $i>0,$ has a finite
set of sons $\{a_{i-1},b_{1},...,b_{p}\}$ with $p>0$,\ as in Lemma 23.

In the first case, we let $U_{z}:=S_{y}$.\ It is a finite leafy tree, possibly
reduced to $z$; its root is $y$. Otherwise we define $U_{z}$ is the union of
the subtrees $S_{a_{i}}$.\ In both cases, $U_{z}\in\mathcal{D}$.

(2) We may have $U_{z}=U_{z^{\prime}}$ for distinct leaves $z,z^{\prime}$.\ We
let $Z$\ contain one and only one leaf $z$ for each $U_{z}$.

(3) In $S=(N_{S},\leq_{S})$, we "contract" as follows the trees $U_{z}$ :

\begin{quote}
for all $z\in Z$, we delete from $S$ the nodes of $U_{z}$, except $z$.
\end{quote}

We obtain a subset $N_{T}$\ of $N_{S}$ and $T:=(N_{T},\leq_{T})$ where
$\leq_{T}$ is the restriction of $\leq_{S}$\ to $N_{T}$.

\bigskip

With these definitions and notation:

\bigskip

\textbf{Lemma 30 }: If $S=(N_{S},\leq_{S})$ is a leafy join-tree then $T$ is
also a leafy join-tree.\ We have $L_{T}=Z$.\ The join mapping $\sqcup_{T}$\ is
the restriction of $\sqcup_{S}$ to $N_{T}$. Furthermore $S=T[U_{z}/z;z\in Z]$.

\textbf{Proof }: These facts are clear.\ We use here the extension (2) of the
basic definition of substitution.\ We have an equality $S=T[U_{z}/z;z\in Z]$,
not only an isomorphism. \hfill$\square$

\bigskip

We denote $T$ by $Contr(S)$. If $S\in\mathcal{D}$, then $Contr(S)$ is a single
node. If $Contr(S)$ is a \emph{star}, \emph{i.e.}, a rooted tree of height 1
(that may be infinite) then, $S\in\mathcal{D}^{+}$.

We denote by $\mathcal{C}[\mathcal{D}]$ the class of join-trees $S$\ in
$\mathcal{T}$\ such that $Contr(S)\in\mathcal{C}.$

\bigskip

\textbf{Theorem 31\ : }If a subclass $\mathcal{C}$ of $\mathcal{T}$\ is
CMSO-constructible, then so is the class $\mathcal{C}[\mathcal{D}]$.

\textbf{Proof }: We will use the definition scheme $\Delta=(\chi,\delta
_{1},\delta_{2},(\theta_{i,j})_{i,j\in\lbrack2]})$ with parameters $X_{1}%
X_{2},X_{3},X_{4}$ for the class $\mathcal{D}$ defined in Theorem 24. We will
combine it with a definition scheme for $\mathcal{C}$ of the form
$\Delta^{\mathcal{C}}=(\chi^{\mathcal{C}},\delta_{1}^{\mathcal{C}}%
,....,\delta_{k}^{C},(\theta_{i,j}^{\mathcal{C}})_{i,j\in\lbrack k]})$ with
parameters $Y_{1},...,Y_{p}$ and copying constant $k$, and get a definition
scheme $\Theta=(\chi^{\prime},\delta_{0}^{\prime},...,\delta_{k}^{\prime
},(\theta_{i,j}^{\prime})_{i,j\in\lbrack0,k]})$ for $\mathcal{C}[\mathcal{D}]$
whose parameters are $X_{1}X_{2},X_{3},X_{4},$ $Y_{1},...,Y_{p},Z$ and copying
constant is $k+1$.

We consider a join-tree $S$\ given by its leaf structure $(L,R)$. We use
notions from Definition 29(2). Our first objective is to identify the
sub-join-trees $U_{z}$ of $S$\ to be contracted.\ 

\emph{Claim} : CMSO\ formulas can express in $(L,R)$ for $x,y$ and $z$ in$\ L$
that $x\sqcup_{S}y$ is a finite node and that $x\ $is a leaf of $U_{z}.$

\emph{Proof} : Clear because $x\sqcup_{S}y$ is finite if and only if the set
$\{z\in L\mid Rzxy\}$ is finite. Furthermore $x$ is a leaf of $U_{z}$ if and
only if $x\sqcup_{S}z$ is finite. \hfill$\square$

\bigskip

Hence, we obtain the leaf structure $(L_{U_{z}},R[L_{U_{z}}])$ of $U_{z}.$ The
leaf structure of $T$ is then $(Z,R[Z])$, where $Z$\ is specified in
Definition 29(2).

We first express that $S\in$ $\mathcal{C}[\mathcal{D}]$ and we also specify
the appropriate parameters.

A formula $\chi^{\prime}(X_{1}X_{2},X_{3},X_{4},Y_{1},...,Y_{p},Z)$\ can
express in $(L,R)$ the following conditions :

\begin{quote}
(i) Properties A1-A5 hold, so that $(L,R)$ is the leaf structure of a
join-tree $S$ in $\mathcal{T}$,

(ii) the parameter $Z$ is as required in Definition 29(2),

(iii) for each $z\in Z$, $(L,R)\models\chi\lbrack L_{U_{z}}](X_{1}X_{2}%
,X_{3},X_{4})$,

(iv) $Y_{1},...,Y_{p}\subseteq Z$ and

(v) $(L,R)\models\chi^{\mathcal{C}}[Z](Y_{1},...,Y_{p})$, which implies that
$Contr(S)\in\mathcal{C}$.
\end{quote}

Next, we describe $N_{S}$\ in terms of these parameters. Without loss of
generality, we assume that $\Delta^{\mathcal{C}}$\ is built so that it
produces from any appropriate leaf structure $(L^{\prime},R^{\prime})$ a
join-tree whose set of nodes is

\begin{quote}
$L^{\prime}\times\{1\}\cup D_{2}\times\{2\}\cup...\cup D_{k}\times\{k\},$
\end{quote}

its set of leaves being $L^{\prime}\times\{1\}.$

\begin{quote}
The set of nodes of the join-tree $\widehat{\Theta}(L,R)$ (isomorphic to $S)$
is defined as

$L\times\{1\}\cup I\times\{0\}\cup D_{2}\times\{2\}\cup...\cup D_{k}%
\times\{k\}$ where:

$L\times\{1\}$ is the set of leaves of $S$, hence the union of the pairwise
disjoint sets of leaves of the $U_{z}$'s,

$I\times\{0\}$ is the union of the sets of internal nodes of the pairwise
disjoint $U_{z}$'s,

$N_{Contr(S)}=Z\times\{1\}\cup D_{2}\times\{2\}\cup...\cup D_{k}\times\{k\},$

$Z\times\{1\}$ is the set of leaves of $Contr(S).$
\end{quote}

The sets $D_{2}\times\{2\}\cup...\cup D_{k}\times\{k\}$ are defined by
$\Delta^{\mathcal{C}}$\ applied to $(Z,R[Z]),$ that is the leaf structure of
$Contr(S)$.

The set of nodes of each $U_{z}$ is $L_{U_{z}}\times\{1\}\cup I_{U_{z}}%
\times\{2\}$ where $I_{U_{z}}$ is defined from parameters $X_{1}\cap L_{U_{z}%
},...,X_{4}\cap L_{U_{z}}\subseteq L_{U_{z}}$ by $\Delta$\ applied to
$(L_{U_{z}},R[L_{U_{z}}])$. From this description, we can (we could) write the
formulas $\delta_{0}^{\prime},...,\delta_{k}^{\prime}.$

It remains to construct the formulas $\theta_{i,j}^{\prime}$ intended to
formalize the order relation $\leq$\ of $S$. We describe it as follows in
terms of the associated strict partial order:

\begin{quote}
$x<y$ \ if and only if,

either $x,y\in N_{U_{z}}$ for some $z\in Z$\ and $x<_{U_{z}}y$,

or $x\in N_{U_{z}}$ for some $z\in Z$, $y$ is internal in $Contr(S)$ and
$z<_{Contr(S)}y$,

or $x,y\in N_{Contr(S)}$ and $x<_{Contr(S)}y$.
\end{quote}

We omit an explicit writing of the formulas. \hfill$\square$

\bigskip

\textbf{Remark 32} : If $\mathcal{C}$ and $\mathcal{E}$ are subclasses of
$\mathcal{T}$, one can defined $\mathcal{C}[\mathcal{E}]$ as the class of
trees $T[U_{z}/z;z\in L_{T}]$ such that $T$ is in $\mathcal{C}$ and each
$U_{z}$ is in $\mathcal{E}$.\ Assuming that $\mathcal{C}$ and $\mathcal{E}$
are CMSO-constructible, we would like to prove that $\mathcal{C}[\mathcal{E}]$
is so by a proof similar to that of Theorem 31.\ Given the leaf structure
$(L,R)$ of some $S$ potentialy in $\mathcal{C}[\mathcal{E}]$, we would need a
CMSO\ identification of the subtrees $U_{z}$ of $S$.\ We had it for the proofs
of Theorems 26 and 31.

\bigskip

Our next construction is based on identifying subtrees of O-trees.

\textbf{Definition 33} : \emph{Subtrees of O-trees}

(1) We recall from Definition 1 that in an O-tree $T=(N,\leq),$ the covering
relation $x<_{c}y$ means that $x$ is a son of $y$.\ We let $\approx$ be the
equivalence relation generated by $<_{c}$.\ We have $x\approx y$ if and only
if there is $z$ such that $x<_{c}^{\ast}z$ and $y<_{c}^{\ast}z$ where *
denotes the reflexive ant transitive closure of $<_{c}$. If $A$ is a subset of
$N$ such that any two elements are $\approx$-equivalent, then $T[A]:=(A,\leq
_{A})$ where $\leq_{A}$ is the restriction of $\leq$ to $A$ is a
\emph{subtree} of $T$ (a "real tree", not a sub-O-tree).\ It may have no root
and/or no leaf.

(2) An equivalence class $C$ is called a \emph{connected component}, and so is
called the subtree $T[C]$ of $T$ that it induces. A connected O-tree, having a
single component, is thus a tree. A connected component may be reduced to a
single node.

(3) We let $T/_{\approx}$ be the quotient structure $(N/_{\approx}%
,\leq_{\approx})$ where $N/_{\approx}$ is the set of connected components and
$C\leq_{\approx}D$ if and only if $x<_{c}y$ for some $x\in C$ and $y\in D$. It
is an "O-tree of trees".

\bigskip

\textbf{Proposition 34}\ : If $T$ is an O-tree, then $T/_{\approx}$ and its
connected components are O-trees.\ If it is a join-tree, so are $T/_{\approx}$
and its connected components. \hfill$\square$

\bigskip

The proof is easy.\ The following examples show that $T/_{\approx}$ and its
connected components may have no leaves, even if $T$ is a leafy join-tree.

\bigskip

\textbf{Examples 35}: \ Let $A=(N_{A},\leq_{A})$ be a leafy connected
join-tree without root.\ For example $N_{A}=\mathbb{N}\times\{0,1\}-\{(0,1)\}$
and $(i,0)\leq_{A}(j,0)$ and $(i,1)<_{A}(j,0)$ where $i\leq j$. Let $a$ be a
fixed internal node.

1) We define a leafy join-tree $T$ with set of nodes $N_{A}\times\mathbb{N}$
ordered as follows, for all $i,j$ in $\mathbb{N}$ and $x,y\in N_{A}$:

\begin{quote}
$(x,i)\leq(y,i)$ if and only if $x\leq_{A}y$

$(x,i)<(y,j)$ where $i\neq j$ holds if and only if $j<i$ and $a\leq_{A}y$.
\end{quote}

Each $N_{A}\times\mathbb{\{}i\mathbb{\}}$ is a connected component. The
quotient tree$\ T/_{\approx}\simeq(\mathbb{N},\geq)$ has root 0 and no leaf.
Each node has a son, \ and $T/_{\approx}$ has a unique connected component.

2) We now define the leafy join-tree $S:=(N_{S},\leq_{S})$ where
$N_{S}:=\mathbb{N}\cup(N_{A}\times\mathbb{N})$ and the following partial
order, for all $i,j$ in $\mathbb{N}$ and $x,y\in N_{A}$:

\begin{quote}
$i\leq_{S}j$ if and only if $j\leq i$,

$(x,i)<_{S}j$ if and only if $j\leq i$,

$(x,i)<_{S}(y,i)$ if and only if $x\leq_{A}y$.
\end{quote}

Its connected components are $\mathbb{N}$\ and the sets $N_{A}\times\{i\}.$
The subtree $S[\mathbb{N}]=(N_{S},\leq_{S})$ has no leaf. \ \hfill$\square$

\bigskip

\textbf{Theorem 36}\ : Let $T$ be a leafy join-tree such that

(1) its connected components are all in a CMSO-constructible class
$\mathcal{C}$\ and

(2) their leaves are leaves of $T$.\ 

Then $T$ is CMSO-constructible.

\textbf{Proof} : Similar to the previous proofs.\ \hfill$\square$

\bigskip

By Proposition 25 the class $\mathcal{C}$\ \ may be the union of finitely many
CMSO-constructible classes.

\textbf{Open problem 37 }: Exhibit a join-tree in $\mathcal{T}$\ (perhaps $T$
of Example 2) that provably cannot be constructed from its leaf structure by
any CMSO-transduction.

\bigskip

\textbf{Definition 38:} \emph{Alternative definition of leaf structures.}

Leafy finite rooted trees can also be described as families of nonoverlapping
sets\footnote{This description is used in \ \cite{Boj,Cam+}.}. Let
$\mathcal{W}:=(L,\mathcal{M})$ where $L$ is a finite set and $\mathcal{M}$ is
a set of subsets of $L$ satisfying the following conditions:

\begin{quote}
$L\in\mathcal{M}$ and $\{x\}\in\mathcal{M}$ for each $x\in L$,

if $X,Y\in\mathcal{M}$ and $X\cap Y$ is not empty, then $X\subseteq Y$\ or
$Y\subseteq X$ (the sets in $\mathcal{M}$ are not \emph{overlapping}).
\end{quote}

Then $(\mathcal{M},\subseteq)$ is a finite leafy rooted tree $T(\mathcal{W})$
with root $L$ and set of leaves the singletons $\{x\}$ for $x\in L$.
Conversely, every finite leafy rooted tree $T=(N_{T},\leq)$ yields a pair
$\mathcal{W}=(L_{T},\mathcal{M})$ where $\mathcal{M}:=\{N_{\leq x}\cap
L_{T}\mid x\in N_{T}\}$ and $T\simeq T(\mathcal{W})$.\ 

We make $\mathcal{W}$ into the logical structure $(L,M(.))$ where $M(.)$ is
the set predicate such that $M(X)$ holds if and only if $X\in\mathcal{M}%
$.\ \ In some cases, this set predicate can be defined by an MSO\ formula over
a relational structure, for example $(L,edg)$ if $\mathcal{M}$ is the set of
\emph{strong modules}\ of a graph $G$ defined by this structure where
$L=V_{G}$, see\ \cite{Cam+,CouDel}; in this case, $T(L,\mathcal{M})$ is the
\emph{modular decomposition} of $G$

For MSO logic, the descriptions of $T=T(\mathcal{W})$ by $(L_{T},R_{T})$ and
by $(L_{T},M(.))$ are equivalent: $R_{T}$\ can be defined as the set of
triples $(x,y,z)\in L_{T}\times L_{T}\times L_{T}$\ such that $x\in X$ where
$X$\ is the inclusion-minimal set such that $M(X)$ holds and that contains $y$
and $z$. Conversely, a subset $X$ of $L_{T}$ is in $\mathcal{M}$\ if and only
if it is $\{x\in L_{T}\mid R_{T}(x,y,z)$ holds\} for some $y,z\in L_{T}$.

It follows that every CMSO\ formula $\varphi(X_{1},...,X_{p})$ over the leaf
structure $(L_{T},R_{T})$ of a join-tree can be rewritten into an equivalent
one $\psi(X_{1},...,X_{p})$ over the corresponding "set" structure
$(L_{T},M(.))$, and vice-versa. This facts extend to infinite join-trees (we
omit the details).\ The constructions of CMSO\ transductions based on
structures $(L,R)$ and $(L,M(.))$ are thus equivalent, as they are convertible
into one another.

\section{Quasi-trees\ and their substructures}

Quasi-trees have been defined in \cite{CouRwd}.\ The motivation was to define
the rank-width of a countable graph.\ They have been further studied in
\cite{Cou17,Cou21,Cou22}. The notation $\neq xyz...$ means that $x,y$,$z...$
are pairwise distinct.

\bigskip

\textbf{Definitions 39}\ \cite{CouRwd,Cou21,Cou22}: \emph{Betweenness in
trees}

(a) A tree $T$ is an undirected connected graph without loops, parallel edges
and cycles. It has no root.\ It may be (countably) infinite.\ It is defined as
$T=(N,edg)$ where $N$ is the set of nodes and $edg$ is the binary edge
relation.\ Any two nodes are linked by a path (by connectedness) that is
actually unique.

The \emph{degree} of a node is the number of incident edges.\ A \emph{leaf} is
a node of degree 1.\ A tree is \emph{subcubic} if each node has degree at most 3.\ 

(b) The \emph{betweenness} of a tree $T$ is the ternary relation $B$ such that
$Bxyz$ if and only if $\neq xyz$ and $y$ is on the path between $x$ and $z$.

(c) The betweenness of a rooted tree is that of the associated unrooted
undirected tree. \hfill$\square$

\bigskip

For shortening formulas relative to ternary structures $(N,B)$, we will denote
$Bxyz\vee Byxz\vee Bxzy$ by $Axyz$ to mean that $x,y$ and $z$ are
\emph{aligned} on a same "path". Furthemore $B^{+}x_{1}...x_{n}$, \ for
$n\geq4$, will stand for the conjunction

\begin{quote}
$\ Bx_{1}x_{2}x_{3}\wedge Bx_{2}x_{3}x_{4}\wedge...\wedge Bx_{n-2}x_{n-1}%
x_{n}$
\end{quote}

and $[x,y]$ will denote the set $\{x,y\}\cup\{u\in N\mid Bxuy\}$, called an
\emph{interval} or a "path".

\bigskip

Betweenness in partial orders has been studied in\ \cite{Cou20} and in
previous articles cited there.

\bigskip

\textbf{Proposition 40}\ (Proposition 9 of \cite{CouRwd}) : For every tree
$T$, the following properties hold for all nodes $x,y,z$ and $u$ :\ 

B1: $Bxyz\Longrightarrow\ \neq xyz,$

B2: $Bxyz\Longrightarrow Bzyx,$

B3: $Bxyz\Longrightarrow\lnot Bxzy,$

B4:\ $Bxyz\wedge Byzu\Longrightarrow Bxyu\wedge Bxzu,$

B5: $Bxyz\wedge Bxuy\Longrightarrow Bxuz\wedge Buyz,$

B6: $Bxyz\wedge Bxuz\Longrightarrow y=u\vee B^{+}xyuz\vee B^{+}xuyz,$

B7: $\neq xyz\Longrightarrow Axyz\vee\exists w.(Bxwy\wedge Bywz\wedge Bxwz).$
\hfill$\square$

\bigskip

\textbf{Remarks 41 }: (1) Except B7, these properties are universal. Property
B7\ asserts the existence of a node $w$ if $\neq xyz\wedge\lnot Axyz$.\ The
unicity of such $w$ is a consequence of the universally quantified B1-B6 (see
Lemma 11 of\ \cite{CouRwd})\ in arbitrary ternary structures.\ We will denote
$w$ by $M(x,y,z)$. Using the notation $M(x,y,z)$ will mean that we have $\neq
xyz\wedge\lnot Axyz.$

(2) It is empty if and only if $T$ has at most two two nodes, and then, B1-B7
hold trivially.

\bigskip

\textbf{Definition 42 } \cite{CouRwd} : \emph{Quasi-trees and related notions}.

A \emph{quasi-tree} is a pair $Q=(N,B)$ where $N$ is the set of \emph{nodes}
and $B$\ is a ternary relation that satisfies Properties B1-B7 (for all
nodes).\ We say that $x$ is \emph{internal} if $Byxz$ holds for some nodes $y$
and $z$; otherwise, $x$ is a \emph{leaf\footnote{This definition applies to
every ternary structure $(N,B)$.}}.

We let $\mathcal{Q}$ be the set of \emph{leafy} quasi-trees\ $Q=(N,B)$,\emph{
i.e.}, those such that every internal node is $M(x,y,z)$ for some leaves
$x,y,z$.\ We cannot have $\left\vert N\right\vert =3$.\ We accept $\left\vert
N\right\vert \leq2$ and $B=\emptyset$ as there are no internal nodes and thus,
each node is a leaf.

\bigskip

\textbf{Remarks 43}: (1) With B4 that we will always assume in our proofs
concerning ternary structures $(N,B)$, $B^{+}x_{1}...x_{n}$ implies
$Bx_{i}x_{j}x_{k}$ for all $1\leq i<j<k\leq n.$

(2) For leaves $x,y,z,$ we never have $Bxyz$, hence properties B1-B6\ are
vacuously true and B7\ reduces to:

B7': $\neq xyz\Longrightarrow\exists w.(Bxwy\wedge Bywz\wedge Bxwz).$

(3) In quasi-trees, we consider an infinite interval $[x,y]$ between nodes $x$
and $y\neq x$ as a "path".\ It is lineraly ordered in such a way that $x\leq
u\leq y$ for all $u\in\lbrack x,y]$ by $u<v$ if and only if $(u=x\wedge
v=y)\vee$ $(\neq xuv\wedge Bxuv)\vee(\neq uvy\wedge Buvy).$

(4) An example of a quasi-tree is $([0,1],B)$ where $[0,1]$ is the closed
interval of rational numbers between 0 and 1, and $Bxyz$ holds if and only if
$x<y<z\vee z<y<x.$ There is an infinite "path" between its two leaves 0 and 1.
This quasi-tree is not leafy. If we delete 0 and 1, we obtain a quasi-tree
without leaves.

(5) Lemma 2.7 of\ \cite{Cou21} shows that B1-B7 imply the following:

B8: $\neq xyzu\wedge Bxyz\wedge\lnot Ayzu\Longrightarrow Bxyu.$

This a universal property, hence it is invariant under taking
substructures.\ The induced substructures of quasi-trees, we call them
\emph{partial quasi-trees}, are the models of B1-B6 and B8 by Theorem 3.1
of\ \cite{Cou21}. We will study them below in a logical perspective.

(6) If a ternary structure $(N,B)$ of cardinality at least 3 satisfies
Property B7, then its Gaifman graph is connected.\ This connectedness property
is not implied by B1-B6 that are universal.

\bigskip

The next proposition shows that a quasi-tree can be seen as the undirected
version of a join-tree.

\bigskip

\textbf{Proposition 44} : If $Q=(N,B)\in\mathcal{Q}$ and $r\in N$, then
$T(Q,r):=(N,\leq_{r})$ is a join-tree with root $r$ whose ancestor ordering is:

\begin{quote}
$x\leq_{r}y$ $:\Longleftrightarrow x=y\vee y=r\vee Bxyr$.\ 
\end{quote}

Its join operation is defined by $x\sqcup_{r}r:=r$, $x\sqcup_{r}y:=x$ if $x=y$
and $x\sqcup_{r}y:=M(x,y,r)$ if $\neq rxy.$ Conversely, every rooted join-tree
in $\mathcal{T}$\ is of this form.

\textbf{Proof: }Lemma 14\ of \cite{CouRwd} proves the first
assertion.\ Conversely, if $T=(N,\leq)$ is a rooted join-tree with root $r$,
we let $B\ $be defined as follows:

\begin{quote}
$Bxyz:\Longleftrightarrow(x<y<z)\vee(z<y<x)\vee(x\neq z\wedge x\sqcup z<y).$
\end{quote}

It is easy to see that $\leq$ is $\leq_{r}$, defined from $B$\ as above.

The hypothesis that $T$ is a join-tree implies that $\left\vert N\right\vert
\geq3$.\ $\ $ \hfill$\square$

\bigskip

In our proofs, we will use B2 without explicit mention.

\bigskip

\textbf{Proposition 45 : }Let $x,y,z$ be three leaves of a quasi-tree $(N,B)$,
$a:=M(x,y,z)$ and $u$ be any node.\ 

(i) We have $Buax\vee(Buay\wedge Buaz).$

(ii) The internal node $a$ is $M(u,x,y)$, $M(u,y,z)$ or $M(u,x,z)$.

(iii) If $Buax$ holds, then $a$ is $M(u,x,y)$ or $M(u,x,z)$.

\textbf{Proof :}\ (i) Assume that $Buax$ does not hold.\ If $Baux$ holds, we
have $B^{+}xuay$ and $B^{+}xuaz$.\ Hence $Buay$ and $Buaz$. Otherwise,
$b:=M(x,a,u)$ is defined (because $x$ is a leaf leaves, so that $Baxu$ can
hold). We have $Bxba$ and $Bxay$, hence we have $B^{+}xbay$ by B5, whence
$Bbay$.\ As $Buba$ holds, B4\ gives $B^{+}ubay$. Hence $Buay$ holds. The proof
is similar for $z$.

(ii) We have two cases.\ If $Buax$ does not hold, then $Buay\wedge Buaz$ holds
and yields $a=M(u,y,z)$.\ If $Buax$ hold and so does $Buay$, we have
$a=M(u,x,y)$.\ If $Buay$ does not hold, we have $a=M(u,x,z)$ by the very first case.

(iii) Let $Buax$ hold.\ If $Buay$ holds, we have $a=M(u,x,y)$.\ Otherwise we
have $Buax\wedge Buaz$ by (i).\ Hence, $a=M(u,x,z)$.\ $\ $ \hfill$\square$

\bigskip

We recall that $[x,y]$ denotes the \emph{interval} $\{x,y\}\cup\{u\in N\mid
Bxuy\}$.

\bigskip

\textbf{Proposition 46 : }Let $x,y,z,u$ be pairwise distinct leaves of a
quasi-tree $Q=(N,B)$. Let $a:=M(x,y,z)$ and $b:=M(x,u,z).$

We have the following cases.

(1) $\left\vert [x,y]\cap\lbrack z,u]\right\vert =1$.$\ $Then $a=b,[x,y]\cap
\lbrack z,u]=[x,z]\cap\lbrack y,u]=[x,u]\cap\lbrack y,z]=\{a\},$
$M(x,y,u)=M(y,z,u)=a,$ and $Bvaw$ holds for any two distinct leaves $v,w$ in
$\{x,y,z,u\}.$

(2) $[x,y]\cap\lbrack z,u]=\emptyset$.\ Then $a\neq b$, $[x,z]\cap\lbrack
y,u]=[x,u]\cap\lbrack y,z]=[a,b],$ $B^{+}xabz$,$B^{+}xabu$, $B^{+}yabz$ and
$B^{+}yabu$ hold, $M(x,y,u)=a$ and $M(y,z,u)=b$.

(3) $\left\vert [x,y]\cap\lbrack z,u]\right\vert \geq2$.\ Then either
$[x,z]\cap\lbrack y,u]=\emptyset$ or $[x,u]\cap\lbrack y,z]=\emptyset$; the
consequences follow from Case (2) by appropriate substitutions of
variables.\hfill$\square$

\bigskip

We will express Case (1) by $Exyzu$ where $E$ is a 4-ary relation, and
similarly Case (2) by $Sxyzu.$

\bigskip

\textbf{Proof: }(1) \ Let $T(Q,z)=(N,\leq_{z})$ be the join-tree with root $z$
constructed by Proposition 44. Then $x<_{z}a,b<_{z}z$.

If $\left\vert [x,y]\cap\lbrack z,u]\right\vert =1$ then $a=b$, the
consequences are clear.\ See Figure 4(a).

(2) We consider again $T(Q,z)$. If$\ [x,y]\cap\lbrack z,u]=\emptyset$, then
$a<_{z}b$.\ The consequences are clear.\ See Figure 4(b).

(3) If $\left\vert [x,y]\cap\lbrack z,u]\right\vert \geq2$, we consider
$T(Q,y)$. If $a<_{y}x\sqcup_{y}u$, then $[x,z]\cap\lbrack y,u]=\emptyset$.\ If
$a>_{y}x\sqcup_{y}u$, then $[x,u]\cap\lbrack y,z]=\emptyset$.\ See Figures
4(c) and 4(d) respectively.\ \hfill$\square$%

\begin{figure}
[ptb]
\begin{center}
\includegraphics[
height=2.3462in,
width=3.5544in
]{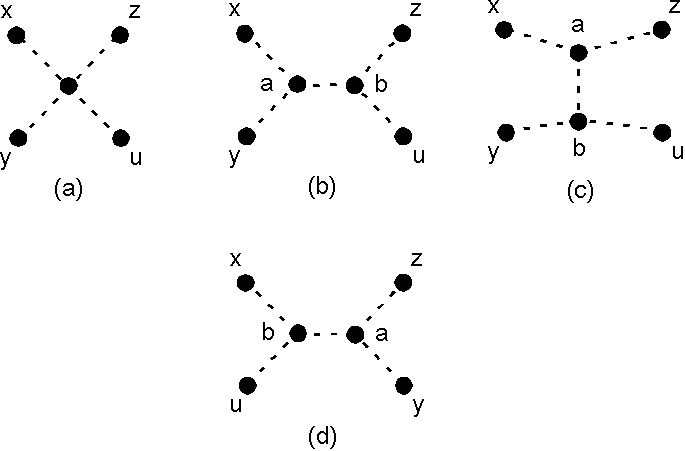}%
\caption{For Remark 47}%
\end{center}
\end{figure}

\bigskip

\textbf{Remarks 47 :}\ (1) Figure 4\ shows the four possible relative
configurations of four leaves $x,y,z$ and $u$. The dotted lines represent "paths".\ 

(2) $M(x,y,z)=M(x,y,u)=M(x,z,u)\ $implies $M(x,y,z)=M(y,z,u)$\ and
$[xy]\cap\lbrack zu]=\{M(x,y,z)\}.$ This follows from the definitions, we are
in Case (1). However, $M(x,y,z)=M(x,y,u)$ does not imply $M(x,y,z)=M(x,z,u).$
See Figure 4(b).

(3) It is not true that $u\neq v\wedge Exyzu\wedge Exyzv\Longrightarrow
Exyuv.$ Take for an example $a:=M(x,y,z),Babu$ and $Babv$.\ We have then
$b=M(a,u,v)\neq a$. See Figure 5.

(4) The relations $E_{Q}$ and $S_{Q}$ can be determined from the restriction
of $B_{Q}$\ to $L_{Q}\times I_{Q}\times L_{Q}$, because from this restriction,
one can determine the intersections $[x,y]\cap\lbrack z,u]$, $[x,z]\cap\lbrack
y,u]$ and $[x,u]\cap\lbrack y,z]$.\ \hfill$\square$%

\begin{figure}
[ptb]
\begin{center}
\includegraphics[
height=1.4373in,
width=1.4667in
]{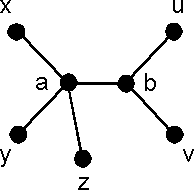}%
\caption{See Remark 47(3)}%
\end{center}
\end{figure}

\subsection{Leafy quasi-trees.}

By Definition 42, a leafy\ quasi-tree $Q$ is the union of the intervals
$[x,y]_{Q}$ for all leaves $x,y.$ If $X$\ is a set of leaves of a quasi-tree
$Q=(N,B)$ and $Y:=X\cup\{M(x,y,z)\mid x,y,z\in X\}$, then $(Y,B[Y])$ is a
leafy quasi-tree: for proving this, just consider the rooted join-tree
$T(Q,u)$ for some $u$ in $X$, cf.\ Proposition 44.\ Then the remark in
Definition 1(c) shows that $Y$ ordered by $\leq_{u\text{ }}$ is a join-tree,
which yields the result.\ We omit details.

\bigskip

Our aim is now to describe a leafy quasi-tree $Q$ from an appropriate relation
on its set of leaves $L$, similarly to what we did for join-trees by means of
the ternary relation $R$ in Theorem 9. The 4-ary relations $E$ and $S$ are
defined after Proposition 42.\ We call $S$\ the \emph{separation relation} of
$Q$, and $(L,S)$ its \emph{separation structure}. The next proposition is an
immediate corollary of Proposition 42. The symbol $\oplus$\ denotes the
exclusive or.

\bigskip

\textbf{Proposition 48} : The following properties hold for all leaves
$x,y,z,u$ of a quasi-tree.\ 

\begin{quote}
S1: $Sxyzu\vee Exyzu\Longrightarrow\ \neq xyzu.$

S2: $Sxyzu\Longleftrightarrow Szuxy\Longleftrightarrow Syxzu.$

S3: $Exyzu\Longleftrightarrow Eyxzu\Longleftrightarrow Eyzux.$

S4: $\neq xyzu\Longrightarrow Exyzu\oplus Sxyzu\oplus Sxzyu\oplus
Sxuyz.$\ \hfill$\square$\ 
\end{quote}

The following properties follow from S1 and S4.

\begin{quote}
S'4: $Exyzu\Longleftrightarrow\ \neq xyzu\wedge\lnot Sxyzu\wedge\lnot
Sxzyu\wedge\lnot Sxuyz$

S\textquotedblright4: $Sxyzu\Longrightarrow\ \neq xyzu\wedge\lnot
Sxzyu\wedge\lnot Sxuyz.$
\end{quote}

Property S3 shows that $E$ is stable by every permution of its four
arguments.\ Property S4 is illustrated in Figure 4. By S'4 the relation $E$ is
definable from $S$ and we will use it to shorten formulas. All our results
will be formulated in terms of 4-ary structures $(L,S)$.\ By Properties S2 and
S4, $S$ is stable by less permutations of its arguments than $E$.

Remark that Property S4 is similar to property A10 (cf.\ Proposition 6).

\bigskip

\textbf{Example 49}: Figure 5\ shows a finite quasi-tree for which we have:

\begin{quote}
$a=M(x,y,z)=M(x,y,u)=M(x,z,u).$

$b=M(x,u,v)=M(y,u,v)\neq a.$

$Sxyuv\wedge Sxzuv\wedge Exyzu\wedge\lnot Eyzuv.$ \ \ \ \hfill$\square$\ 
\end{quote}

The following propoposition is formally similar to Proposition 4. We will
denote by $EA$ where $A$ is a set of cardinality at least 4, the property that
$Exyzu$ holds for any four distinct $x,y,z,u$ in $A$. We will denote by
$SAB$\ where $A,B$ are disjoint sets of cardinality at least 2 the property
that $Sxyzu$ holds for any distinct $x,y$ in $A$ and $z,u$ in $B$.

\bigskip

\textbf{Lemma 50:} Let $Q$ be a leafy quasi-tree with set of internal nodes
$I_{Q}$. Let $x,y$ be distinct leaves and $a,b,c$ be distinct internal nodes.

(i) $Bxay$ holds if and only if $a=M(x,u,y)$ for some leaf $u$.

(ii) $Bxab$ holds if and only if $a=M(x,u,v)$ and $b=M(x,v,w)$ for some leaves
$u,v$ and $w$ such that $Sxuvw$ holds.

(iii) $Babc$ holds if and only if $Buab,Bubc$ and $Buac$ hold for some leaf
$u$.

\textbf{Proof: }(i) Assume $Bxay.$ By Proposition 45, we have $a=M(x,u,v)$ for
some leaves $u,v$.\ If $Buay$ holds, then $a=M(x,u,y)$. If not, we have $Byax$
and $Byav$ hence $a=M(x,v,y)$. Conversely, if $a=M(x,u,y)$ then $Bxay$ holds.\ 

(ii) Assume $Bxab$.\ We have $b=M(x,v,w)$ for some leaves $v,w$ by Proposition
45.\ Hence we have $B^{+}xabv$, and so $Bxav$.\ Then $a=M(x,u,v)$ for some
leaf $u$ by (i).\ Clearly, $Sxuvw$ holds.\ 

Conversely, let $a=M(x,u,v)$ and $b=M(x,v,w)$ for some leaves $u,v$ and $w$
such that $Sxuvw$ holds. We have $Bxav$ and $Bxbv$. We cannot have $a=b$
since\ $Exuvw$ does not hold.\ Hence, we have $B^{+}xabv$ or $B^{+}%
xbav$.\ Because of $Sxuvw$ the latter cannot hold.\ Hence we have $B^{+}xabv$
and so $Bxab$.

(iii) Assume $Babc$.\ Then $a=M(u,v,c)$ for some leaves $u,v$ by Proposition
45.\ Hence, we have $Buac$ and also $B^{+}uabc$ and so we have $Buab,Bubc$ and
$Buac$. Conversely, if $Buab,Bubc$ and $Buac$ hold, then $B^{+}uabc$ holds by
a now routine argument, and so does $Babc$. \ \hfill$\square$\ 

\bigskip

\textbf{Proposition 51 : } A\ leafy quasi-tree $Q=(N,B)$ in $\mathcal{Q}$\ is
characterized by the structure $(L_{Q},S_{Q})$.

\textbf{Proof:} If $Q$ has exactly three leaves, then $S_{Q}$ is empty, but
since $Q$\ is leafy, it has a unique internal node and is characterized by
$(L_{Q},S_{Q})$. Leaving this case, we will reconstruct $Q$ \emph{u.t.i.} from
$(L,S)=(L_{Q},S_{Q})$ by using Propositions 45 and Lemma 50.\ 

We define $j:L\times L\times L\rightarrow N$ by $j(x,y,z):=M(x,y,z)$ provided
$x,y$ and $z$ are pairwise distinct. We denote by $m,p,q$ such triples.\ 

We denote by $xyz$ a triple of leaves $(x,y,z)$.\ 

We have $j(m)=j(p)$ where $m=xyz$ and $p=uvw$ if and only if

\begin{quote}
$(x,y,z)$ is a permutation of $(u,v,w)$ or $E\{x,y,z,u,v,w\}$ holds,
cf.\ Proposition 46.\hfill\ (1)
\end{quote}

We write then $m\equiv p$. This equivalence relation is definable from $E$,
hence from $S$\ via S3. For distinct leaves $x$ and $y$, and pairwise
nonequivalent triples $m,p,q$ as above, we have, by Lemma 50:

\begin{quote}
$B(x,j(m),y)\Longleftrightarrow m\equiv xzy$ for some leaf $z$,\ \hfill\ (2)

$B(x,j(m),j(p))\Longleftrightarrow m\equiv xuv$, $p\equiv$ $xvw$ and $Sxuvw$
holds for some leaves $u,v$ and $w$, \hfill(3)

$B(j(m),j(p),j(q))\Longleftrightarrow B(u,j(m),j(p)),$ $B(u,j(p),j(q))$\ and
$B(u,j(m),$ $j(q))$\ \ hold for some leaf $u$. \hfill(4)
\end{quote}

If $R=(N,B^{\prime})\in\mathcal{Q}$\ is such that $(L_{R},S_{R})=(L_{Q}%
,S_{Q}),$ if $j$ is as above and $j^{\prime}$ is defined similarly for $R$,
then we have identity between the sets of leaves of $R$ and $Q$ and
furthermore, $j(m)=j(p)$ if and only if $j^{\prime}(m)=j^{\prime}(p)$ because
$S_{R}=S_{Q}$ and the equalities $j(m)=j(p)$ and $j^{\prime}(m)=j^{\prime}(p)$
are defined in the same ways from $S_{Q}$ and $S_{R}.$ We obtain a bijection
between the internal nodes of $R$ and $Q$.

By (2)-(4) above, the betweenness relations of $R$ and $Q$\ are defined in the
same ways from $S_{Q}$ and $S_{R}.$ Hence, $R=(N_{R},B_{R})$ and\ $Q=(N_{Q}%
,B_{Q})$ are isomorphic.\hfill\ $\square$

\bigskip

In a separation structure $(L,S)$, the relation $S$ is empty if $\left\vert
L\right\vert \leq3$ or if $Exyzu$ holds for any four leaves $x,y,z,u$.\ In the
latter case, the quasi-tree has a single internal node. Proposition 51 is
valid.\ We admit quasi-trees $(N,\emptyset)$ where $\left\vert N\right\vert
\leq2.$

\bigskip

\textbf{Corollary 52}\ : Let $Q$ be a leafy quasi-tree with set of internal
nodes $I_{Q}$.\ Its relations $S_{Q},E_{Q}$ and $B_{Q}$\ are FO definable in
the the relational structure $(N_{Q},B_{Q}\cap(L_{Q}\times I_{Q}\times
L_{Q}))$.

\textbf{Proof} : Note that $M_{Q}$ is FO definable in $(N_{Q},B_{Q}\cap
(L_{Q}\times I_{Q}\times L_{Q}))$.\ The statements follow from the definitions
Remark 47(5) and Proposition 50.\ \hfill$\square$

\bigskip

Our next aim is to find an FO\ axiomatization of the separation relations that
characterize leafy quasi-trees from their leaves.

\bigskip

\textbf{Proposition 53}\ : The class of separation structures of leafy
quasi-trees is hereditary.

\textbf{Proof }: Let $X\subseteq L_{Q}$ where $Q=(N,B)$ is a quasi-tree.\ Let
$Y:=X\cup\{M_{Q}(x,y,z)\mid x,y,z\in X\}$.\ We get a leafy quasi-tree
(cf.\ the beginning of Section 3.1) $P=(Y,B[Y]),$ possibly reduced to one or
two nodes : in this case $B[Y]$\ is empty.

We claim that $(X,S_{Q}[X])=(L_{P},S_{P})$.

We use Proposition 46.\ Note that $[x,y]_{P}\cap\lbrack z,u]_{P}$
$\subseteq\lbrack x,y]_{Q}\cap\lbrack z,u]_{Q}$.\ If $S_{Q}xyzu$ holds where
$x,y,z,u\in X$, then $[x,y]_{Q}\cap\lbrack z,u]_{Q}$ is empty and so is
$[x,y]_{P}\cap\lbrack z,u]_{P}$ .\ Hence, we have $S_{P}xyzu.$ Conversely, if
we have $S_{P}xyzu$, then $[x,y]_{P}\cap\lbrack z,u]_{P}$ contains the two
distinct nodes $M_{P}(x,y,z)$ and $M_{P}(x,z,u)$, so does $[x,y]_{Q}%
\cap\lbrack z,u]_{Q}$ hence\ $S_{Q}xyzu$ holds.\hfill$\square$

\bigskip

\textbf{Construction 54\ }: \emph{A leafy quasi-tree constructed from its
leaves.}

Let $(L,S$) be a 4-ary structure satisfying Properties S1-S4 (where the
relation $E$ is FO-defined from $S$\ by Property S'4) and Property $EQ$\ to be
defined below.\ We construct as follows a ternary structure $(N,B)$ :

\begin{quote}
$N:=P/\equiv$ where $P\subseteq L\times L\times L$ and $\equiv$ is an
equivalence relation on $P,$

$B:=B^{\prime}/\equiv$ \ where $B^{\prime}$ is a ternary relation on $P$ that
is stable under $\equiv$, which means that if $m\equiv m^{\prime},p\equiv
p^{\prime}$ and $q\equiv q^{\prime}$ then $B^{\prime}(m,p,q)$ implies
$B^{\prime}(m^{\prime},p^{\prime},q^{\prime}).$

Furthermore, $P,\equiv$ and $B^{\prime}$ are FO definable in $(L,S)$.
\end{quote}

For this purpose, we let $P$ be the set of triples in $L\times L\times L$
(denoted by $xyz$) whose components are all equal or pairwise distinct.\ 

We define $xyz\equiv uvw$ if and only if either $(x,y,z)$ is a permutation of
$(u,v,w)$ or $E\{x,y,z,u,v,w\}$ holds (the relation $E$ extended to finite
sets of fixed size is FO-definable from $S$ by S'4). This property is
expressible by a quantifier-free formula $\eta(x,y,z,u,v,w).$ We let $EQ$\ be
the uFO sentence expressing the transitivity of $\equiv$, hence that it is
actually an equivalence relation. We only consider structures $(L,S)$ that
satisfy S1-S4 and $EQ$.

We define $B^{\prime}(m,p,q)$ where $m=x_{1}x_{2}x_{3},p=y_{1}y_{2}y_{3}$ and
$q=z_{1}z_{2}z_{3}\ $as follows:

\begin{quote}
$m,p,q\in P$, they are pairwise not equivalent, $\neq y_{1}y_{2}y_{3}$ and

$B^{\prime}(m,p,q):\Longleftrightarrow B_{0}^{\prime}(m,p,q)\vee B_{0}%
^{\prime}(q,p,m)\vee B_{1}^{\prime}(m,p,q)$ where

(a) $B_{0}^{\prime}(m,p,q)$ holds if and only if one has :

(i) either $x_{1}=x_{2}=x_{3},z_{1}=z_{2}=z_{3}$ and $p\equiv x_{1}uz_{1}$ for
some $u$ in $\{y_{1},y_{2},y_{3}\},$

(ii) or $x_{1}=x_{2}=x_{3},\neq z_{1}z_{2}z_{3}$, $p\equiv x_{1}uv$, $q\equiv
x_{1}vw$ and $Sx_{1}uvw$ holds for some $u$ in $\{y_{1},y_{2},y_{3}\}$ and
disctinct $v,w$ in $\{z_{1},z_{2},z_{3}\}$.

(b) $B_{1}^{\prime}(m,p,q)$ holds if and only if, for some $u\in L$, we have:

$B_{0}^{\prime}(uuu,m,p)$ $\wedge$ $B_{0}^{\prime}(uuu,p,q)$ $\wedge
B_{0}^{\prime}(uuu,m,q)$.
\end{quote}

This definition is based on Cases (2)-(4) of Proposition 51. From its
definition, Property $B^{\prime}$\ is stable under $\equiv$.\ An FO\ formula
$\beta(x_{1},x_{2},x_{3},y_{1},y_{2},y_{3},$ $z_{1},z_{2},z_{3})$\ can be
written to express that $B^{\prime}(x_{1}x_{2}x_{3},y_{1}y_{2}y_{3},z_{1}%
z_{2}z_{3})$ holds in a 4-ary structure $(L,S)$.\hfill$\square$

\bigskip

\textbf{Proposition 55\ }: For every FO\ sentence $\varphi$\ over a ternary
structure $(N,B)$, one can construct an FO\ sentence $\widehat{\varphi}$\ such
that, if $(N,B)$ is defined by Construction 54\ from $(L,S)$, then
$(L,S)\models\widehat{\varphi}$\ if and only if $(N,B)\models\varphi$.

\textbf{Proof }: The construction uses induction on the structure of
subformulas $\psi(x,y,...)$ of $\varphi$ with free variables $x,y,$
\emph{etc.} to transform them into $\widehat{\psi}(x_{1},x_{2},x_{3},$
$y_{1},y_{2},y_{3},...).$ The main steps are as follows.

\begin{quote}
$\exists x.\psi(x,y,...)$ becomes $\exists x_{1},x_{2},x_{3}.\pi(x_{1}%
,x_{2},x_{3})\wedge$ $\widehat{\psi}(x_{1},x_{2},x_{3},$ $y_{1},y_{2}%
,y_{3},...)$\ where $\pi$\ defines $P,$

$x=y$ becomes $\eta(x_{1},x_{2},x_{3},y_{1},y_{2},y_{3})$ (we recall that
$\eta$ defines $\equiv),$

$B(x,y,z)$ becomes $\beta(x_{1},x_{2},x_{3},y_{1},y_{2},y_{3},z_{1}%
,z_{2},z_{3})$ where $\beta$ defines $B^{\prime}$.
\end{quote}

Also $\widehat{\lnot\psi}$ is $\lnot\widehat{\psi}$\ and $\widehat
{\theta\wedge\psi}$ is $\widehat{\theta}\wedge\widehat{\psi}.$\ \ We omit
further details.\ \ \hfill$\square$

\bigskip

\textbf{Theorem 56\ }: The class of separation structures of leafy quasi-trees
is first-order definable by a sentence $\Phi.$

\textbf{Proof: }A 4-ary structure $(L,S)$ is the separation structure of a
leafy quasi-tree if and only if it satisfies Properties S1-S5, the FO sentence
$EQ$ and an FO sentence that we will define.

We let $P$ consists of the triples $(x,y,z)$ in $L\times L\times L$ (written
$xyz$) such that either $x=y=z$ or $\neq xyz$. We identify $xxx$ to the leaf
$x$.\ Construction 54\ defines a transformation of $(L,S)$ into $(N,B)$.

By Definition 42, there is a FO sentence $\varphi$ expressing that a ternary
structure $(N,B)$ is a leafy quasi-tree.\ Proposition 51, yields a
FO\ sentence $\widehat{\varphi}$\ expressing in $(L,S)$ that $(N,B)\models
\varphi$, \emph{i.e.}, that $(N,B)$ is a leafy quasi-tree.

It remains to express that $(L,S)$ is the separation structure of
$(N,B)$.\ This is possible by a FO-sentence $\theta$ because $B$ is defined in
$(L,S)$ by $\beta$. Hence $\Phi$\ is the conjunction of S1-S4, $EQ$ and
$\widehat{\varphi}\wedge$ $\theta$. \ \ \hfill$\square$

\bigskip

The sentence $\Phi$ is cumbersome to write and tells nothing about the
relations $S$ and $E$\ in quasi-trees for which we only know presently
Properties S1-S4 by Proposition 48.

The theorem by Los and Tarski (see Hodges \cite{Hod}, Theorem 6.5.4) says that
if a class of finite and infinite relational structures is hereditary and
axiomatizable by a single FO sentence, then it is also by a single uFO
sentence.\ This is the case of separation structures by Proposition 56.

\bigskip

\textbf{Open problem} \textbf{57}\ : \emph{Can one find a readable
FO\ axiomatization of the separation structures of quasi-trees}, similar to
that of Theorem 9\ for the leaf structures of join-trees, to that of
Definition 42 for leafy quasi-trees and to that of Remark 43(5) for partial quasi-trees.

Here is an example of an axiom that one can add to S1-S4 (see Figure 5):

\begin{quote}
S5: $Exyzu\wedge Sxyuv\wedge z\neq v\Longrightarrow\ Sxzuv.$\ 
\end{quote}

\emph{Proof sketch} of its validity: let $a:=M(x,y,u)$ and $b:=M(x,u,v);$ from
the hypotheses, we have $B^{+}xabu$, $Bzau$, hence $B^{+}zabu$.\ With
$B^{+}xabv$, we get $Sxzuv.$\ \hfill$\square$

Property S5\ is not implied by S1-S4.\ We may have S1-S4\ together with
$Exyzu\wedge Sxyuv\wedge z\neq v\wedge\lnot Sxzuv\wedge Exzuv:$ observe that
S1-S4 only concern 4-tuples $xyzu$ and never\ pairs of 4-tuples like
$xyzu$\ and $xyuv$.\ \hfill$\square$

\bigskip

Next, we look for MSO\ transductions that transform a structure $(L,S)$
satisfying $\Phi$\ into an associated leafy quasi-tree.\ 

\bigskip

\textbf{Theorem 58} : There exists a $\leq$-MSO transduction that transforms a
4-ary structure $(L,S)$ satisfying $\Phi$\ into a leafy quasi-tree $Q$ such
that $(L_{Q},S_{Q})\simeq(L,S)$.\ A CMSO-transduction can do that if
$L,$\ whence $Q,$ is finite.

\textbf{Proof: }We first consider a leafy quasi-tree $Q=(N,B)$ and one of its
leaves $r$.\ Let $T(Q,r)$ be the rooted join-tree $(N,\leq_{r})$, cf.
Proposition 44.\ Note that $r$ is not $x\sqcup_{r}y$ for any leaves $x,y$ of
$T(Q,r)$. Let $T^{\prime}(Q,r):=(N-\{r\},\leq_{r}^{\prime})$ where $\leq
_{r}^{\prime}$ is the restriction of $\leq_{r}$\ to $N-\{r\}.$ Then,
$T^{\prime}(Q,r)$ is a leafy join-tree. It may have no root. Its set of leaves
is $L_{Q}-\{r\}.$

\bigskip

\emph{Claim} : The ternary relation $R^{\prime}:=R_{T^{\prime}(Q,r)}$ is FO
definable from $S_{Q}$:

\begin{quote}
for all $x,y$ and $z$ in $L_{Q}-\{r\}$, $R^{\prime}xyz$ holds if and only if

either $x=y=z$ or $x=y\neq z$ or $x=z\neq y$ or $E_{Q}xyzr$ or $S_{Q}xyzr$ or
$S_{Q}xzyz$.\ \hfill$\square$
\end{quote}

Figure 6 illustrates the cases $E_{Q}xyzr$ and $S_{Q}xyzr$ of this claim.%

\begin{figure}
[ptb]
\begin{center}
\includegraphics[
height=1.7936in,
width=2.7146in
]{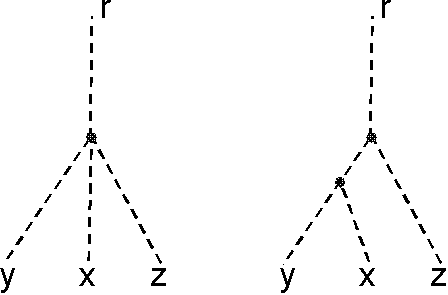}%
\caption{See Theorem 58.}%
\end{center}
\end{figure}

\bigskip

We now start from $(L,S)$ satisfying $\Phi$, hence that it is the separation
structure of a unique (\emph{u.t.i.}) quasi-tree. An MSO-transduction can
transform the separation structure $(L,S)$ of a leafy join-tree into a leaf
structure ($L-\{r\},R^{\prime})$ of a leafy join-tree $T^{\prime
}(Q,r)=(N-\{r\},\leq_{r}^{\prime})$ by means of the claim. Then, an $\leq$
-MSO\ transduction (with copying constant 2) can transform $(L-\{r\},R^{\prime
})$ into $(N-\{r\},\leq_{r}^{\prime})$\ by the result of \cite{CouDel}
recalled in Theorem 15.\ An MSO-transduction (with copying constant 2) can
transform ($N-\{r\},\leq_{r}^{\prime})$ into the join-tree $(N,\leq_{r}),$\ by
adding $r$ to the domain and because:

\begin{quote}
$x\leq_{r}y$ if and only if $x\leq_{r}^{\prime}y\vee y=r\vee x=y=r.$
\end{quote}

Finally, an FO transduction (cf.\ Section 2.1) can transform $(N,\leq_{r})$
into the quasi-tree $Q=(N,B)$ because $B$\ is FO\ definable from $\leq_{r}.$
The composition of these four transductions yields an $\leq$
-MSO\ transduction\ as claimed.\ 

If $L$ is finite, so is $N$.\ The construction of Theorem 18\ makes it
possible to replace auxiliary linear orders by counting modulo 3 set
predicates.\ \hfill$\square$

The choice of $r$ is irrelevant because for any other leaf $r^{\prime}$, the
structure $(L-\{r\},R^{\prime})$ and the corresponding one $(L-\{r^{\prime
}\},R^{\prime\prime})$

are exchangeable by MSO\ transductions.

\subsection{Rank-width and separation structures}

Definitions and background results come from \cite{CouRwd,CouOum,Oum}. The
rank-width of finite graphs is an important notion in the theory of
\emph{Fixed Parameter Tractability}, see \cite{CouEng,Oum}. Graphs are
undirected and \emph{simple}, that is, without loops and parallel edges.

\bigskip

\textbf{Definition 59\ :} \emph{Rank-width of finite graphs}

(a) A \emph{layout} of a finite graph $G=(V_{G},edg_{G})$ is a finite subcubic
tree $T=(N_{T},edg_{T})$ such that $V_{G}$ is the set of its leaves. A tree is
\emph{cubic} (resp. \emph{subcubic}) if its nodes are of degree 1 or 3 (resp.
1,2 or 3).

(b) A \emph{cut} of $T$ is a bipartition of $N_{T}$\ into sets $C_{x}$ and
$C_{y}$ associated with an edge $x-y$, where $C_{x}$ is the set of nodes $z$
such that $z=x$ or $Bzxy$ holds.\ Hence, $C_{x}$ induces a subtree
$(C_{x},edg_{T}[C_{x}])$ containing $x$, and of course, similarly for $C_{y}$.

(c) Let $M_{G}[V,V]$ be the symmetric square adjacency matrix of $G$ over the
field $GF(2)$ such that\ $M_{G}[x,y]=1$ if and only if there is an edge\ $x-y$.

(d) The \emph{rank-width of} $G$ \emph{relative to} $T$, denoted by
$rwd(G,T),$ is the maximal rank (over $GF(2)$) of the rectangular submatrices
$M_{G}[V\cap C_{x},V\cap C_{y}]$ for all edges $x-y$ of $T$. The
\emph{rank-width of} $G$, denoted by $rwd(G),$ is the minimal rank-width
$rwd(G,T)$ for all layouts $T$ of $G$. \hfill$\square$

\bigskip

For each $k$, one can decide in cubic time if a graph $G$\ has rank-width at
most $k$ \cite{Oum}, Theorems 9.5 and 12.8. Furthermore, a CMSO\ sentence can
express that a graph $G=(V_{G},edg_{G})$ has rank-width at most $k$
(\cite{CouOum}, Corollary 2.6 and Section 6.4).\ However, such a sentence does
not yield a CMSO\ transduction that could construct a layout witnessing that
$G$ has rank-width $\leq k$ because it is based of checking the absence of
"forbidden" configurations called \emph{vertex-minors} (see the survey
\cite{Oum}) from a finite set.

\bigskip

For defining the rank-width of a countable graph, quasi-trees offer the
appropriate notion of layout because in this way, the rank-width of a
countable graph is the least-upper bound of those of its finite induced
subgraphs \cite{CouRwd}.

\bigskip

We first consider finite layouts described by their separation structures:
they permit to express the property $rwd(G,T)=k$ by a CMSO\ effectively
constructible sentence.

\bigskip

\textbf{Lemma 60 }: Let $T$ be a finite leafy quasi-tree (based on a tree).\ 

(1) For pairwise distinct leaves $u,v,w,s$, we have:

\begin{quote}
$S_{T}uvws$ if and only if $u,v\in C_{x}$ and $w,s\in C_{y}$ for some edge
$x-y$ of $T$.
\end{quote}

(2) The tree $T$ is cubic (equivalently subcubic) if and only if
$E_{T}=\emptyset.$

\textbf{Proof :} (1)\ If $u,v\in C_{x}$ and $w,s\in C_{y}$, then
$[u,v]\subseteq C_{x}$ and $[w,s]\subseteq C_{y}$, and $C_{x}$ and $C_{y}$ are
disjoint, hence $S_{T}uvws$ holds. Conversely, if $[u,v]\cap\lbrack
w,s]=\emptyset$ then\ Proposition 46(2) shows that $M(u,v,w)\neq M(u,w,s]$;
any edge $x-y$ on the path from $M(u,v,w)$ to $M(u,w,s)$ gives $u,v\in C_{x}$
and $w,s\in C_{y}$ (or vice-versa).

(2) A leafy subcubic quasi-tree is cubic, as it cannot have nodes of degree
2\footnote{See \cite{CouRwd} for detailed definitions.}. If $E_{T}uvws$
holds\footnote{See Property 42(1)\ \ and Figure 4(a).}, then $M(u,v,w)$ has
degree at least 4.\ Conversely, if $x$ has adjacent nodes $u^{\prime
},v^{\prime},w^{\prime},s^{\prime}$, then there are leaves $u,v,w,s$ ,
respectively in $C_{u^{\prime}},C_{v^{\prime}},C_{w^{\prime}},C_{s^{\prime}}$
(defined from $x-u^{\prime}$, $x-v^{\prime}$ etc.), and disjoint paths between
$x$ and these leaves. Hence, $E_{T}uvws$ holds. \ \hfill$\square$

\bigskip

We recall (Example 49) that for disjoint sets of leaves $A$ and $B$, $S_{T}AB$
means that $S_{T}uvws$ holds for any two distinct $u,v\in A$ and $w,s\in B.$

\bigskip

\textbf{Lemma 61 }: If $T$ is a layout of a finite graph $G$, then $rwd(G,T)$
is the maximal rank of the matrix $M_{G}[A,B]$ for all sets $A,B\subseteq
L_{T}=V_{G}$ such that $S_{T}AB$\ holds.

\textbf{Proof }: The previous lemma shows that, if $A\subseteq C_{x}$ and
$B\subseteq C_{y}$ for some edge $x-y$, then $S_{T}AB$ holds.\ Conversely,
assume we have $S_{T}AB$\ for some subsets of $V_{G}$. If follows from the
proof of\ Lemma\ 59\ below that $A\subseteq C_{x}$ and $B\subseteq C_{y}$ for
some edge $x-y$.\ We omit details. \hfill$\square$

\bigskip

We now extend these definitions and results to infinite graphs. For finite
graphs, the following definitions coincide with the previous ones. A leafy
quasi-tree $Q$ is \emph{cubic} if its relation $E_{Q}$ is empty. This
definition is equivalent to Definition 18 of \cite{CouRwd}.

\bigskip

\textbf{Definition 62\ \cite{CouRwd} :} \emph{The rank-width of countable
graphs}

(a) A \emph{layout} of a graph $G=(V_{G},edg_{G})$ is a cubic and leafy
quasi-tree $Q=(N_{Q},B_{Q})$ such that $V_{G}$ is its set of leaves.

(b) A set $C\subseteq N_{Q}$ is \emph{convex} if $[u,v]\subseteq C$ whenever
$u,v\in C$. A \emph{cut} of $Q$ is a bipartition of $N_{Q}$ into convex sets
$C$ and $D$ having at least two nodes.\ Hence, $S_{Q}CD$\ holds because
$[u,v]\cap\lbrack w,s]=\emptyset$ if $u,v\in C$ and $w,s\in D$.

(c) The \emph{rank-width of} $G$ \emph{relative to} $Q$, denoted by $rwd(G,Q)$
is the maximal rank of a rectangular matrix $M_{G}[V\cap C,V\cap D]$ for a cut
$\{C,D\}$ of $Q$. If $C$ or $D$ is singleton, the corresponding rank is 1 or
0.\ We can neglect such cuts as the rank of a graph with at least one edge is
at least 1.

(d) The \emph{rank-width of} $G$, denoted by $rwd(G)$ is the minimal
rank-width $rwd(G,Q)$ for a layout $Q$ of $G$. It may be infinite. We might
accept as layout a quasi-tree that is subcubic or not leafy.\ The resulting
notion of rank-width is the same, see \textbf{\ \cite{CouRwd}.}

\bigskip

\textbf{Lemma 63 }: Let $A,B$\ be disjoint sets of leaves of a leafy
quasi-tree $Q=(N,B)$.\ If $S_{Q}AB$\ holds, then $A\subseteq C$\ and
$B\subseteq D$ for some cut $\{C,D\}.$

\textbf{Proof:} Let $u$\ be a leaf of a leafy quasi-tree $Q=(N,B)$ and
$T:=(N,\leq)$ be the join-tree $T(Q,u)$ with root $u$ defined in Proposition
44.\ Let $x,y,z$ be other leaves of $Q$: we have $x\sqcup y<x\sqcup z$ if and
only if $S_{Q}xyzu$ holds. This fact follows from Proposition 46\ since
$x\sqcup_{T}y=M_{Q}(x,y,u)$ and $x\sqcup z=M_{Q}(x,z,u)$.

Let $A,B$\ be disjoint sets of leaves of $Q$ as above, such that $S_{Q}AB$
holds.\ Let $u\in B$\ and $T=(N,\leq)$ be the join-tree $T(Q,u)$. We define
$C$ and $D$.\ 

Let $x\in A$. For each $y\in A$, we have $x\leq x\sqcup y<u$.\ Let $C:=\{w\in
N\mid w\leq x\sqcup y$ for some $y\in A$\}. It is convex because for any two
$w,w^{\prime}$ in $C$, we have $w\sqcup w^{\prime}\leq x\sqcup y$ for some
$y\in A,$ hence all nodes in $[w,w^{\prime}]_{Q}$ are below $w\sqcup
w^{\prime}$, hence below $x\sqcup y,$ whence in $C$.

We prove by contradiction that $B\cap C$ is empty.\ If $z\in B\cap C,$ we have
$z\leq x\sqcup y$ for some $x,y\in A$. But we have $S_{Q}xyzu$ which implies
that $a:=M(x,y,u)=x\sqcup y<b:=M(x,z,u)=x\sqcup z$.\ We cannot have $z\leq
x\sqcup y$.

Consider now $D:=N-C$.\ We have $B\subseteq D$. Let $w$ in $D$ that is not a
leaf.\ There is in $B$\ a leaf $z<w$.\ As $C$ is downwards closed, the
interval $[z,u]_{Q}$ is contained in $D$ and contains $[w,u]_{Q}$.\ For
another $w^{\prime}$ in $D$ we have similarly $z^{\prime}<w^{\prime}$ for some
leaf $z^{\prime}$ in $B$.\ It follows that the interval $[w,w^{\prime}]_{Q}$
is contained in $[w,t]_{Q}\cup\lbrack t,w^{\prime}]_{Q}$ where $t:=w\sqcup
w^{\prime}$, hence $D$ is convex.\hfill$\square$

\bigskip

\textbf{Theorem 64\ } \cite{CouRwd}, Theorem 23 : The rank-width of a graph is
the least upper-bound of those of its finite induced subgraphs.

\bigskip

The following proposition makes no use of forbidden vertex-minors.

\bigskip

\textbf{Proposition 65\ : }For each $k$, a CMSO sentence can express in a
relational structure $(V,edg,S)$ that $rwd(G,Q)\leq k$ where $G:=(V,edg)$\ and
$(V,S)$ is the separation structure of a layout $Q$ of this graph.

\textbf{Proof :} First we observe that an FO sentence $\varphi$\ can express
that $(V,S)$ is the separation structure of a layout $Q$\ of $G$, where
$V=L_{Q}$.\ This follows from Theorem 56\ and the observation that $Q$ is
subcubic if and only if $E_{Q}$ is empty.

The rank of a matrix $M_{G}[X,Y]$ where $X$ is finite is the cardinality of a
maximal set\footnote{Maximal for inclusion.} $Z\subseteq X$ such that the row
vectors $M_{G}[z,Y]$ for $z\in Z$\ are independent. These vectors are
independent if and only if their sum is not $\overrightarrow{0}$, that is, if
and only if the sum in $GF(2)$ of the values $M_{G}[z,y]$'s is not 0 for some
$y\in Y$.\ A CMSO\ formula $\beta(Z,Y)$ can express this in $(V,edg)$ by means
of the cardinality predicate $C_{2}$. Hence, a CMSO\ formula $\alpha(X,Y,Z)$
can express that $Z\subseteq X$ is such a maximal set, and so, that the rank
of $M_{G}[X,Y]$ is $\left\vert Z\right\vert $.\ 

It follows that, a relational structure $(V,edg,S)$ as above, defines a layout
$Q=(V,S)$ of $G:=(V,edg)$ such that $rwd(G,Q)\leq k$ if and only if it
satisfies $\varphi$ and the following property:

\begin{quote}
for all finite subsets $X,Y,Z$ of $V$, if $S_{Q}XY$ and $\alpha(X,Y,Z)$ hold,
then $\left\vert Z\right\vert \leq k.$
\end{quote}

A CMSO\ sentence $\rho_{k}$ can express these conditions because the
finiteness of a set is CMSO expressible, $\alpha(X,Y,Z)$ is CMSO, $\varphi
$\ and $S_{Q}XY$ are FO, and so is the property $\left\vert Z\right\vert \leq
k$\ for each fixed integer $k$.

This upper-bound may not exist: then $rwd(G,Q)=\omega.$

The sentences $\rho_{k}$ can be replaced by a unique CMSO\ sentence $\rho(Z)$
such that $rwd(G,Q)$ is the maximal cardinality of a finite set $Z$ such that
$(V,edg,S)\models\rho(Z)$. Hence, an alternative formulation is

\begin{quote}
$rwd(G,Q)$ is the maximal cardinality of a finite subset $Z$ of $V$\ such that
$(V,edg,S)\models\rho(Z)$
\end{quote}

where $\rho(Z)$ is the CMSO\ formula expressing that there exist finite
subsets $X,Y$ of $V$ such that $\varphi\wedge\alpha(X,Y,Z)$ and $S_{Q}XY$
hold. We have here a unique CMSO\ expression that does not depend on $k,$ but
uses instead a "maximum cardinality" set predicate, cf.\ \cite{CouDur}%
.\hfill.\hfill$\square$

\bigskip

An open problem consists in finding classes of graphs for which a CMSO\ or a
$\leq$-MSO\ transduction can produce a separation structure yielding
rank-width at most a fixed $k$. The article\textbf{\ \cite{BojGro+} }shows
that from graphs of bounded linear clique-width (equivalently of bounded
linear rank-width), a CMSO\ transduction can build a clique-width term of
width $f(k)$.\ This term can be converted into a layout witnessing bounded
rank-width by an MSO-transduction.

\ 

\subsection{Partial quasi-trees}

\textbf{Definition 66\ : }\emph{Partial quasi-trees\ as Induced Betweenness
Relations in Quasi-Trees.}

Let $Q=(N,B)$ be a quasi-tree and $X\subseteq N$.\ Then $Q[X]:=(X,B[X])$ where
$B[X]$ is the restriction of $B$ to $X,$ is a \emph{partial quasi-tree}.\ It
is not necessarly a quasi-tree because some nodes $M(x,y,z)$ where $x,y,z$
$\in X$ may not be in $X$. Its leaves are defined as for quasi-trees in
Definition 42.\ An internal node of $Q$ may become a leaf in $Q[X]$ but not
vice-versa.\ In \cite{Cou21} we denoted by \textbf{IBQT}\ (for \emph{Induced
Betweenness in Quasi-Trees}) the class of partial quasi-trees. It is
\emph{hereditary} by definition, \emph{i.e.}, it is closed under taking
induced substructures.\ Theorem 3.1\ of \cite{Cou21}\ states that it is
axiomatized by an uFO\ sentence consisting of B1-B6 and the following

\begin{quote}
B8: $\neq xyzu\wedge Bxyz\wedge\lnot Ayzu\Longrightarrow Bxyu.$
\end{quote}

A pair $(N,\emptyset)$ is a "degenerated" partial quasi-tree obtained by
deleting all internal nodes of a leafy quasi-tree.\ It satisfies B1-B6 and B8,
and each node in $N$ is a leaf.

We write $Q^{\prime}\longrightarrow Q$ if $Q^{\prime}$ is a quasi-tree and $Q$
is a partial quasi-tree obtained from $Q$ by deleting some internal nodes.\ It
follows that $L_{Q}=L_{Q^{\prime}}$.\ It is not possible to reconstruct
$Q^{\prime}$\ from $Q$ in a unique way, even if every node of $Q$ belongs to
some triple of $B$. Here is an example.%

\begin{figure}
[ptb]
\begin{center}
\includegraphics[
height=1.2609in,
width=2.2866in
]{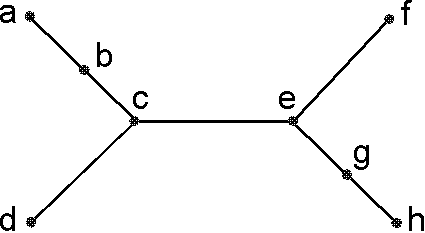}%
\caption{For Example 67.}%
\end{center}
\end{figure}

\bigskip

\textbf{Example 67 }: Let $Q^{\prime}$ be the quasi-tree $(N^{\prime
},B^{\prime})$\ shown (as a finite tree) in Figure 7. If we delete the nodes
$c$ and $e$, we get a partial quasi-tree $Q=(N,B)$ whose nodes are
$a,b,d,f,g,h$ \ and its\ betweenness relation consists of
$Babd,Babf,Bfgh,Bdgh$ and the triples derived from $B^{+}abgh.$ We have
$Q^{\prime}\longrightarrow Q$.

However, we also have $Q^{\prime\prime}\longrightarrow Q$ where $Q^{\prime
\prime}$ is associated with the tree obtained from that of Figure 7 by fusing
$c$ and $e$ (by contracting the edge $c-e$).\ However, we have $S_{Q^{\prime}%
}adfh$ and $E_{Q^{\prime\prime}}adfh$, whence $\lnot S_{Q^{\prime\prime}}%
adfh$.\ The separation relations $S_{Q^{\prime}}$ and $S_{Q^{\prime\prime}}%
$\ (on $L_{Q}$) distinguish $Q^{\prime\prime}$ from $Q^{\prime}$. They keep
the "memory" of the possible quasi-trees from which $Q$\ is derived.
\ \hfill$\square$

\bigskip

\textbf{Theorem 68 }: Let $(N,B,S)$ be a relational structure such that $B$ is
ternary and $S$ is 4-ary.\ There is at most one leafy quasi-tree $Q^{\prime
}=(N^{\prime},B^{\prime})$ such that $Q^{\prime}\longrightarrow Q=(N,B)$,
$L_{Q}=L_{Q^{\prime}}$ and $S_{Q^{\prime}}=S$.\ The existence of such a pair
is FO expressible.\ If it does exist, one can construct $Q^{\prime}$ by an
$\leq$-MSO\ transduction, and by a CMSO\ one if $L$ is finite.

\bigskip

Hence, $Q=(N,B)$ is a partial quasi-tree derived from a unique leafy
quasi-tree $Q^{\prime}$ specified by a 4-ary relation $S$ on $L_{Q}$\ defined
as in Definition 42.\ 

\bigskip

\textbf{Lemma 69}\ : Let $Q=(N,B)$ be a leafy quasi-tree such that $S_{Q}$ is
not empty.\ Each leaf belongs to some 4-tuple of $S_{Q}$.\ 

\textbf{Proof} : Assume we have $S_{Q}xyuv$.\ Let $z$ be a leaf not among
$x,y,u,v$.\ By Property S4, we have $E_{Q}xyzu\vee S_{Q}xyzu\vee S_{Q}xzyu\vee
S_{Q}xuyz.$ The last three cases give the result, and\ $Exyzu\wedge S_{Q}xyuv$
yields $S_{Q}szuv$ by S5\ of Problem 57. \ \hfill\ \hfill\ \ \hfill$\square$

\textbf{Proof of Theorem 68: }Let $(N,B,S)$ and $(Q,Q^{\prime})$ be as in the statement.

\emph{Case 1}\ : $S\neq\emptyset$. By Lemma 69, $L_{Q}=L_{Q^{\prime}}$ is
determined from $S=S_{Q^{\prime}}$.\ It follows from Proposition 51\ that
$Q^{\prime}$ is determined in a unique way. So is $Q$ since $N-L_{Q}$\ is the
set of nodes $y$ such that $Bxyz$ holds for some leaves $x,z$.\ 

\emph{Case 2} : $S=\emptyset,B\neq\emptyset$. Then, $Q^{\prime}$ is a star
with center $r$ and $Q^{\prime}=Q$. Then $B_{Q}$\ contains all triples
$(x,r,y)$ for distinct $x,y$ in $L$.

\emph{Case 3} : $S=\emptyset,B=\emptyset$. Here, $Q^{\prime}$ is a star and
$Q$ is $Q^{\prime}$ minus the center.

The other properties follow from these observations and from Theorems 56\ and
58 for Case 1.\ \ \hfill$\square$

\bigskip

It follows that if $(L,S)$ and $(L,S^{\prime})$ are the leaf structures of two
leafy quasi-trees $Q$ and $Q^{\prime}$ and if $S\subseteq S^{\prime}$, then
$S\subseteq S^{\prime}$ and $Q\simeq Q^{\prime}$. We may have $S=\emptyset$.

\subsection{Topological embeddings}

In order to illustrate the notion of separation of a partial quasi-tree, we
recall from \cite{Cou21}\ a topological characterization of quasi-trees and
their induced substructures. Detailed definitions are in this article.

\bigskip

\textbf{Definition 70} : \emph{Topological embeddings of quasi-trees.}

(1) We let $\mathcal{L}$ be a tree of \emph{half-lines} in the plane.\ Let
$N$\ be a countable subset of the union $\mathcal{L}^{\#}$ of these
half-lines.\ For distinct $x,y,z$ in $N$, we let $B_{\mathcal{L}}xyz$ mean
that $y$ is on the unique path in the topological space $\mathcal{L}^{\#}$,
denoted by $[x,z]_{\mathcal{L}}$, where a \emph{path} is a union of finitely
many line segments homeomorphic to the real interval [0,1]. It is itself
homeomorphic to [0,1]. In the context of a particular embedding, we denote in
the same way a node and the corresponding point in $\mathcal{L}^{\#}$.\ 

Examples of embeddings are shown in Figure 8.\ Part (a) shows that of a
quasi-tree $Q$\ whose nodes are $a,b,c,d,e,f,g,u,v,w$ and $x$.\ Its leaves are
$a,c,d,e$ and $g$.\ It is the quasi-tree of a finite tree with root $u$. Part
(b) shows an induced substructure of it where nodes $u,v,w$ and $x$ have been
deleted (cf. Definition 70). \ \hfill$\square$

\bigskip

Theorem 4.4\ of \cite{Cou21} shows that a structure $(N,B_{\mathcal{L}})$ as
constructed above is a partial quasi-tree and conversely that every partial
quasi-tree is of this form.

\bigskip

\textbf{Lemma 71 :} In a quasi-tree $Q=(N,B_{\mathcal{L}})$ every internal
node $M_{Q}(x,y,z)$ is a point $m$ of $\mathcal{L}^{\#}$ that belongs two or
three half-lines and is the origin of at least one of them.\ 

\textbf{Proof }: If $x,y,z$ would be on a same line, then $M_{Q}(x,y,z)$ would
be undefined.\ Hence, $m$ must belong to two or three half-lines, and be the
origin of at least one of them.\ It may be the origin of two or three of
them.\ \ \hfill$\square$

An example in Figure 8(a) is $v=M_{Q}(a,b,d)$, belonging to three half-lines.

Let $Q=(N,B_{\mathcal{L}})$ be a leafy quasi-tree embedded in a tree
$\mathcal{L}$ of half lines.\ Let $c$ be a point in $\mathcal{L}^{\#}$ not in
$N$. The topological space $\mathcal{L}^{\#}-\{c\}$ consists of two connected
components. Any two nodes $x,y$ are in a same component, written $x\sim_{c}y$
if and only if $c\notin\lbrack x,y]_{\mathcal{L}}.$ The two equivalence
classes are denoted by $X_{c}$ and $Y_{c}$; they form a cut of $Q$. With this notation:

\bigskip%

\begin{figure}
[ptb]
\begin{center}
\includegraphics[
height=2.4232in,
width=1.7582in
]{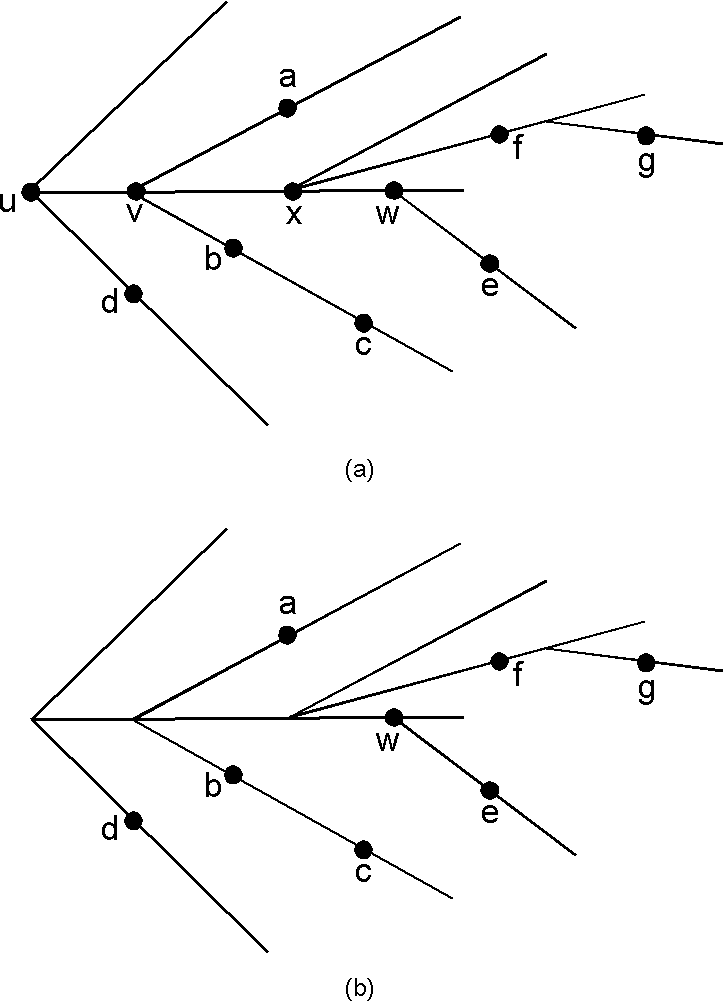}%
\caption{See Definition 70}%
\end{center}
\end{figure}

\bigskip

\textbf{Lemma 72 :} If $X_{c}$ and $Y_{c}$ contain at least two leaves, they
define as follows a 4-ary relation on pairwise distinct leaves:

\begin{quote}
$S_{c}xyzu:\Longleftrightarrow$ $x$ and $y$ are in an equivalence class and
$z$ and $u$ are in the other.
\end{quote}

The separation relation $S$ of $Q$\ is the union of the relations $S_{c}.$

\textbf{Proof: }If\textbf{ }$S_{c}xyzu$ holds, then the path in $\mathcal{L}%
^{\#}$\ between $x$ and $z$ must go through $c$.\ We have $c\in\lbrack
x,z]_{\mathcal{L}}\cap\lbrack y,z]_{\mathcal{L}}\cap\lbrack x,u]_{\mathcal{L}%
}\cap\lbrack y,u]_{\mathcal{L}}$. Since $x$ and $y$ are in the same connected
component of $\mathcal{L}^{\#}-\{c\}$, the path $[x,y]_{\mathcal{L}}$ does not
go through $c$.\ Hence $M(x,y,z)$ belongs to this path and $x\sim
_{c}M(x,y,z)\sim_{c}y.$ Similarly $M(x,z,u)$ is in the other connected
component hence $M(x,y,z)\neq M(x,z,u)$ and we have $Sxyzu$. Note that
$c\in\lbrack M(x,y,z),M(x,z,u)]_{\mathcal{L}}-N.$

Conversely, assume $Sxyzu$.\ There is a point $c$ not in $N$\ on the path
$[M(x,y,z),$\ $M(x,z,u)]_{\mathcal{L}}$\ \ because this path consists of
finitely many segments forming a connectred subset of $\mathcal{L}^{\#}$.\ It
follows that $x,M(x,y,z)$ and $y$ are in one of the connected components of
$\mathcal{L}^{\#}-\{c\}$\ and $z,M(x,y,z)$ and $u$ are in the other.\ Hence we
have $S_{c}xyzu$. \ \hfill$\square$

\bigskip

\textbf{Remark 73} : From an embedding of a quasi-tree $Q$ in a tree of
half-lines $\mathcal{L}$, we get an embedding of any partial quasi-tree
included in $Q$. If $Q$ is leafy and we delete all internal nodes in the
embedding, the induced betweenness relation $B_{Q}[L]$ is empty and gives no
information about the resulting embedding, as $L$ is a set without any
structure.\ However, the separation relation $S_{Q}$ describes, up to
homeomorphism, the part of a topological space $\mathcal{L}^{\#}$ that
represents the relative positions of the nodes $M(x,y,z)$ in the original embedding.

\bigskip

\textbf{Conclusion} : We raised a few open questions.\ The main one is Problem
57, that is, to obtain a readable uFO axiomatization of the separation
structures of leafy quasi-trees.\ 

\bigskip

\textbf{Acknowledgement : }I thank M.\ Bojanczyk for the nice construction of
Theorem 18\ that suggested me several extensions to infinite join-trees and
quasi-trees, and M.\ Kant\'{e} for his comments. I thank M.\ Raskin for help
with the FO\ prover Z3.

\bigskip

\textbf{Disclaimer }: No single sentence of this article has been produced by
ChapGPT or any other similar tool. No dog or cat has suffered from my writing
this paper. Neither did my wife.


\begin{thebibliography}{99}                                                                                               %


\bibitem {Boj}M. Bojanczyk, \emph{The category of MSO transductions}, May
2023, ArXiv 230518039.

\bibitem {BojGro+}M.\ Bojanczyk, M. Grohe, and M. Pilipczuk. Definable
decompositions for graphs of bounded linear cliquewidth. \emph{Log. Methods
Comput. Sci.}, \textbf{17}(1) (2021), Paper No. 5, 40, 2021.

\bibitem {Cam+}R.\ Campbell \emph{et al.}, CMSO-transducing tree-like graph
decompositions, december 2024, ArXiv 2412.04970.

\bibitem {Cha+}M. Changat, P. Narasimha-Shenoi and G. Seethakuttyamma,
Betweenness in graphs: A short survey on shortest and induced path
betweenness. \emph{AKCE Int. J. Graphs Combinat}. \textbf{16} (2019) 96--109.

\bibitem {Chen}Y. Chen and J. Flum, Forbidden induced subgraphs and the
Lo\'{s}-Tarski Theorem. \emph{J. Symb. Log.} \textbf{89 }(2024) 516-548.

\bibitem {Chv}V. Chvatal, Antimatroids, betweenness, convexity, in
\emph{Research Trends in Combinatorial Optimization}, Springer (2008) 57--64.

\bibitem {FPIT}B. Courcelle, Fundamental properties of infinite Trees.
\emph{Theor. Comput. Sci. }\textbf{25} (1983) 95-169.

\bibitem {CouX}B. Courcelle, The monadic second-order logic of graphs X:
Linear orderings. \emph{Theor. Comput. Sci.} \textbf{160} (1996) 87-143

\bibitem {CouXVI}B. Courcelle, The monadic second-order logic of graphs XVI :
Canonical graph decompositions. \emph{Log. Methods Comput. Sci.} \textbf{2}
(2006) Issue 2.

\bibitem {CouRwd}B.\ Courcelle, Several notions of rank-width for countable
graphs. \emph{J. Comb. Theory B} \textbf{123} (2017) 186-214

\bibitem {Cou17}B. Courcelle, Algebraic and logical descriptions of
generalized trees. \emph{Log. Methods Comput. Sci. }\textbf{13} (2017) Issue 3.

\bibitem {Cou20}B. Courcelle, Betweenness of partial orders. \emph{RAIRO
Theor. Informatics Appl. }\textbf{54}, 7 (2020)

\bibitem {Cou21}B.\ Courcelle, Axiomatization of betweenness in
order-theoretic trees. \emph{Log. Methods Comput. Sci.} \textbf{17} (2021)
Issue 1.

\bibitem {Cou22}B. Courcelle, Induced betweenness in order-theoretic trees.
\emph{Discret. Math. Theor. Comput. Sci.} \textbf{23} (2021) Issue 2.

\bibitem {CouDel}B.\ Courcelle and C. Delhomm\'{e}, The modular decomposition
of countable graphs. Definition and construction in monadic second-order
logic. \emph{Theor. Comput. Sci.} \textbf{394 }(2008) 1-38

\bibitem {CouDur}B. Courcelle and I.\ Durand, Computations by fly-automata
beyond monadic second-order logic. \emph{Theor. Comput. Sci.} \textbf{619}
(2016) 32-67

\bibitem {CouEng}B. Courcelle and J. Engelfriet, \emph{Graph structure and
monadic second-order logic, a language theoretic approach}, Cambridge
University Press, 2012.

\bibitem {CouOum}B.\ Courcelle and S. Oum, Vertex-minors, monadic second-order
logic, and a conjecture by Seese. \emph{J. Comb. Theory B} \textbf{97} (2007) 91-126

\bibitem {Fra}R. Fra\"{\i}ss\'{e}, Theory of relations, \emph{Studies in
Logic} \textbf{Vol}. 145, North-Holland, 2000

\bibitem {GanRub}T. Ganzow and S. Rubin, Order-invariant MSO is stronger than
counting MSO in the finite, \emph{STACS\ 2008, 25th Annual Symposium on
Theoretical Aspects of Computer Science, Bordeaux, France},\ \emph{Proceedings
LIPIcs 1, Schloss Dagstuhl - Leibniz-Zentrum f\"{u}r Informatik,} Germany,
2008, pp. 313-324.

\bibitem {Hod}W.\ Hodges, \emph{Model theory}, Cambridge University Press, 1993.

\bibitem {Oum}D. Kim and S. Oum, Vertex-minors of graphs: A survey.
\emph{Discret. Appl. Math.} \textbf{351} (2024) 54-73.

\bibitem {Twd}I. Kriz and R.\ Thomas, Clique-sums, tree-decompositions and
compactness. \emph{Discret. Math.} \textbf{81} (1990) 177-185

\bibitem {Z3}The Z3\ prover, https://www.microsoft.com/en-us/research/project/z3-3/
\end{thebibliography}
\end{document}